\documentclass[preprint,12pt,authoryear]{elsarticle}

\makeatletter
\def\ps@pprintTitle{%
  \let\@oddhead\@empty
  \let\@evenhead\@empty
  \let\@oddfoot\@empty
  \let\@evenfoot\@oddfoot
}
\makeatother

\usepackage{amssymb,mathtools,xcolor,hyperref,cleveref,tcolorbox,adjustbox,multirow,multicol,subfigure}
\graphicspath{{./figures/}}
\usepackage[ruled,linesnumbered]{algorithm2e}
\Crefname{algocf}{Algorithm}{Algorithms}
\usepackage{comment}

\begin{document}

\begin{frontmatter}

\title{Bayesian Data Integration and Model Updating in Probabilistic Digital Twins}

\affiliation[gni_label]{organization={Georg Nemetschek Institute
Artificial Intelligence for the Built World, Technical University of Munich},
            % addressline={Walther-von-Dyck-Straße 10},
            city={Munich},
            % postcode={85748},
            country={Germany}}

\affiliation[deltares_label]{organization={Deltares},
            city={Delft},
            country={The Netherlands}}

\affiliation[tud_label]{organization={Faculty of Civil Engineering and Geosciences, Delft University of Technology},
            city={Delft},
            country={The Netherlands}}         

\affiliation[epfl_label]{organization={School of Architecture, Civil and Environmental Engineering, EPFL},
            city={Lausanne},
            country={Switzerland}}

\affiliation[era_label]{organization={Engineering Risk Analysis Group, Technical University of Munich},
            % addressline={Theresienstr. 90},
            city={Munich},
            % postcode={80333},
            country={Germany}}
            
\affiliation[mdsi_label]{organization={Munich Data Science Institute, Technical University of Munich},
            % addressline={Theresienstr. 90},
            city={Garching},
            % postcode={80333},
            country={Germany}}

\author[gni_label]{Dafydd Cotoarbă\corref{cor}}
\ead{dafydd.cotoarba@tum.de}
\author[deltares_label]{Antonis Mavritsakis}
\author[deltares_label,tud_label]{Timo Schweckendiek}
\author[epfl_label]{Ian FC Smith}
\author[era_label,mdsi_label]{Daniel Straub}

\cortext[cor]{Corresponding author}

\begin{abstract}
This paper proposes a framework for sequential Bayesian updating in probabilistic digital twins using data from multiple sources collected over time. The framework distinguishes between property data and behavior data, according to how observations relate to the digital state. Property data directly inform a subset of the uncertain model parameters and can therefore be incorporated using efficient dedicated Bayesian updating methods. Behavior data provide indirect information through the prediction model and require the solution of a Bayesian inverse problem. To reduce the associated computational cost, an active-learning-based surrogate strategy is introduced that evolves with the digital state and adaptively refines the prediction model in regions relevant to posterior inference. The framework is evaluated using an idealized cantilever beam example and a geotechnical case study of a canal wall subject to varying water levels and corrosion. The examples show how complementary data sources, including both equality observations and inequality information from observed system survival, can inform uncertain parameters and system reliability. The proposed surrogate-based updating scheme yields posterior estimates and prediction accuracy comparable to those of physics-based updating while requiring substantially fewer model evaluations. The proposed framework thus supports multi-source updating and reliability assessment in probabilistic digital twins of engineering systems.
\end{abstract}

\begin{keyword}
probabilistic digital twins \sep geotechnical design and construction \sep uncertainty quantification \sep bayesian updating \sep surrogate modelling \sep reliability analysis

\end{keyword}

\end{frontmatter}

% \linenumbers

\section{Introduction}
\label{sec:introduction}

The growing availability of monitoring data throughout the lifecycle of infrastructure assets increases the interest in methods that systematically integrate observations with computational prediction models. Such methods have been developed in the context of system identification, where indirect measurements are combined with physical models to infer the underlying causes or states of a system \citep{Smith2016}. In structural engineering, Bayesian model updating has been used for several decades to include monitoring data in predictive models \citet{beck1998updating}. Today, Bayesian updating and system identification find widespread application in Structural health monitoring (SHM) and related engineering fields \citep[e.g.,][]{vanik2000bayesian,yuen2006efficient,goulet2013structural,pasquier2015robust,Smith2016,kamariotis2023off}. More recently, these methods have been included in the digital twin (DT) concept, which finds increasing application in architecture, engineering, construction, operations and management as a means to support data-driven decision-making. Despite its potential, several challenges still limit the widespread adoption of DTs in construction practice, including a lack of trust, difficulties in integrating heterogeneous data sources, limited predictive model accuracy, and scalability issues \citep{opoku_barriers_2023,albalkhy_digital_2024}.

In traditional DT implementations, some of these challenges arise from the use of deterministic models and parameters. Achieving reliable predictions under such assumptions requires detailed physical models that accurately capture system behavior. In practice, however, model uncertainty is often substantial and is further compounded by uncertainties from material variability, environmental conditions, and measurement errors. Moreover, deterministic modeling frameworks do not naturally support model updating as new information becomes available. Instead, uncertainty is commonly addressed through safety factors. While safety factors can ensure a safe, conservative design, they cannot be combined with monitoring data that reflect the actual performance of the structures.

The Probabilistic Digital Twin (PDT) addresses these limitations \citep{kapteyn2021probabilistic,chaudhuri2023predictive,torzoni2024digital}. In \citet{cotoarbua2025probabilistic} we propose a PDT for geotechnical design and construction. The PDT framework formalizes the observational method \citep{peck1969advantages,spross2017observational,bismut2023optimal}, which is a common design approach in geotechnical engineering and is incorporated into Eurocode 7 \citep{CEN_04}. It specifies the definition of contingency actions for deviations from acceptable behavior limits during the construction phase. 

The PDT framework provides an explicit representation of uncertainty and uses probabilistic modeling and Bayesian inference to update the system state as new data is observed. A central element of the PDT proposed in \citet{cotoarbua2025probabilistic} is the classification of observational data into two categories, which clarifies how different data sources contribute to the evolution of the digital state. The first category, referred to as \emph{property data}, consists of measurements of material or physical properties and can be used to directly update the random variables corresponding to model parameters. The second category, \emph{behavior data}, comprises data obtained by observing system response or performance, such as displacements or settlements, which can only be used to update model parameters through Bayesian inversion.

Bayesian updating with behavior data is more challenging than with property data, as repeated likelihood evaluations require a prediction model that is often computationally costly (e.g., a finite element model). To address this challenge, a range of sampling-based methods has been proposed, including Markov Chain Monte Carlo (MCMC) \citep{hastings1970monte,gilks_markov_1995,gelman_bayesian_1995,beck_bayesian_2002,Neal_2011}, sequential Monte Carlo methods \citep{doucet2001sequential,del_moral_sequential_2006,cappe2007overview,ching_transitional_2007,betz_transitional_2016,chopin2020introduction,kamariotis2023off},
and Bayesian Updating with Structural reliability methods (BUS) \citep{straub2011reliability,straub2015bayesian,diazdelao_bayesian_2017,betz2018bayesian}.

These methods have also found increasing application in geotechnical engineering, where reliability assessment is often required under large epistemic uncertainties and limited data. High-fidelity geotechnical models are computationally expensive and therefore repeated evaluations within Bayesian inference are challenging. As a result, surrogate modeling approaches have been widely adopted for reliability analysis in geotechnical applications, including sheet pile walls in dyke bodies \citep{van_den_eijnden_metamodelling_2022}, monopile foundations \citep{depina_reliability_2016,liu_adaptive_2022}, slope stability problems \citep{kang_system_2015}, and strip footings \citep{al-bittar_kriging-based_2018,soubra_probabilistic_2019}. To improve the computational efficiency of training a surrogate model for high-dimensional problems, active learning strategies focus model refinement on regions of interest. The integration of active learning with sampling-based Bayesian updating methods, such as BUS, has been extensively studied in the context of reliability analysis \citep[e.g.,][]{wang_highly_2020,song_buak-ais_2022,rossat_bayesian_2022,van_den_eijnden_metamodelling_2022,jiang_adaptive_2023,prentzas_quantified_2024,prentzas_active_2025,huynh_parallel_2025,thompson_comparative_2025,pires_reliability_2025}. 
In this work, we investigate how active learning strategies can be systematically integrated within the PDT framework. By combining surrogate modeling, active learning, and Bayesian updating, both the digital twin and the surrogate model can evolve continuously and in parallel as new data is obtained. This enables accurate and computationally efficient inference over time, supporting the practical implementation of probabilistic digital twins for engineering applications.

The specific objectives of this work are (1) to investigate approaches to Bayesian updating when multiple data sources are available at different time points; (2) to study how approximation errors introduced by Bayesian updating schemes and surrogate models affect the final inference results; and (3) to evaluate the efficiency and accuracy of surrogate models with active learning compared with physics-based models. This contributes to the ultimate goal of systematically integrating continuous data streams from various sources within the PDT framework.

The paper is structured as follows. In \Cref{sec:PDT}, the PDT framework is summarized. \Cref{sec:pdt_updating} presents model updating over multiple time steps and the incorporation of data from multiple sources within the PDT. \Cref{sec:seq_gpr}, proposes a sequential surrogate model. The results for two example problems are then shown in \Cref{sec:example_cantilever,sec:example_canal}. The paper ends with discussions and conclusions in \Cref{sec:concluding_remarks}.

\section{Probabilistic Digital Twin Concept}
\label{sec:PDT}

\begin{figure*}[h!]
\centering
\includegraphics[width=\textwidth]{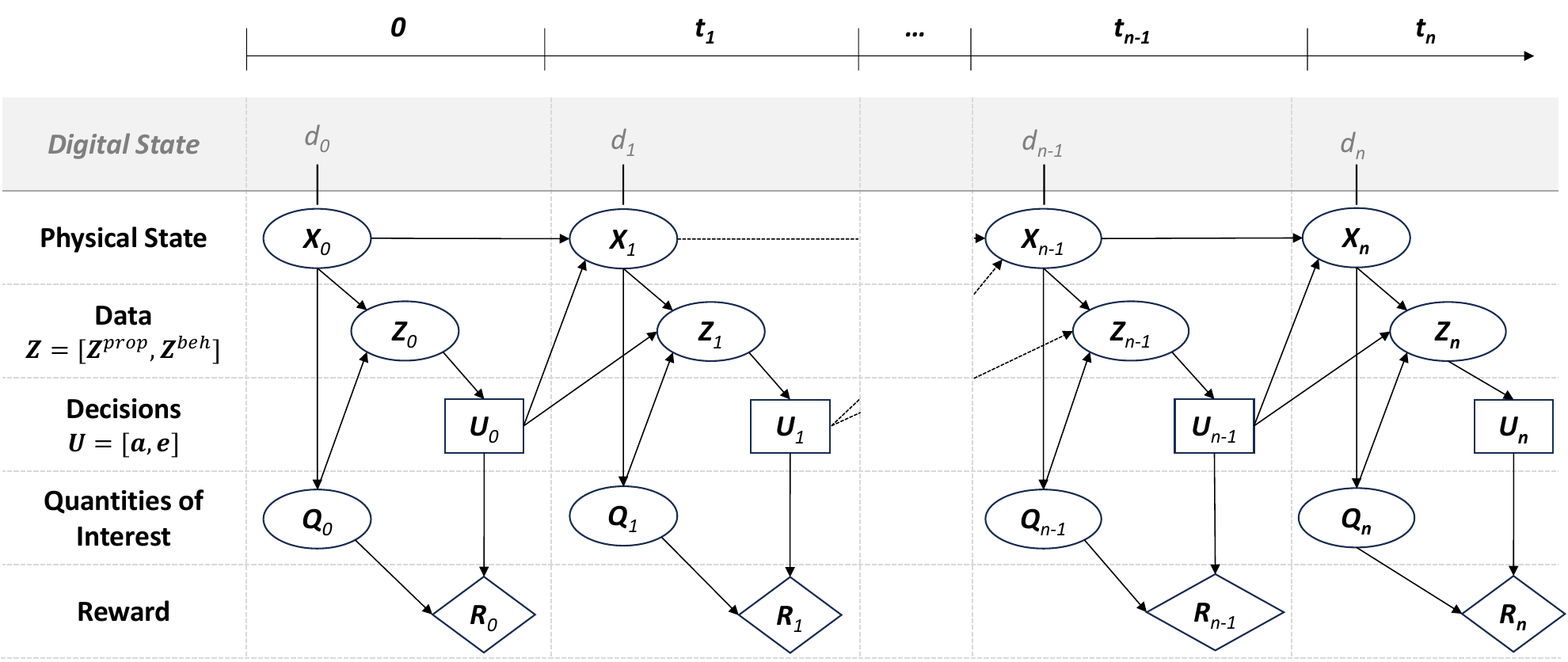}
\caption{The proposed PDT model is represented by an influence diagram to highlight the conditional dependencies of the individual components as shown in \citet{cotoarbua2025probabilistic}.}
\label{fig:dt_if_diagram}
\end{figure*}

A probabilistic digital twin (PDT) for geotechnical design and construction should enable the probabilistic integration of data, behavioral predictions, and decision-making across multiple time steps. \Cref{fig:dt_if_diagram} shows the PDT presented in \citet{cotoarbua2025probabilistic}. It incorporates the Markov assumption and describes the relationships between the physical system state $\mathbf{X}_t$, measurement data $\mathbf{Z}_t = [\mathbf{Z}_{p,t}, \mathbf{Z}_{b,t}]$, quantities of interest $\mathbf{Q}_t$, decisions $\mathbf{U}_t = [\mathbf{a}_t, \mathbf{e}_t]$, and rewards $\mathbf{R}_t$. Decisions are categorized into \textit{controlling actions} $\mathbf{a}_t$, which directly affect the physical system, and \textit{information collection actions} $\mathbf{e}_t$, which aim to reduce epistemic uncertainty by collecting additional data.

In practical engineering applications, the true system state $\mathbf{X}_t$ is not directly observable and is modeled probabilistically. Therefore, the model underlying the PDT is a hidden Markov model \citep{koller2009probabilistic}. As a result, the system knowledge at time $t$ is represented by a probability distribution over $\mathbf{X}_t$, which is here referred to as the \textit{digital state} $d_t$. It is also known as the \textit{belief} in partially observable Markov decision processes (POMDP).

The digital state at time $t$ is defined as $d_t=p \left( \textbf{x}_t|\textbf{z}_{0:t},\textbf{u}_{0:t-1} \right)$. This is the probabilistic distribution of the current system state conditional on all past and current measurement data and all past decisions. As new data are acquired and decisions are made, the digital state evolves over time, resulting in a continuously updated digital twin. Following \citet{kapteyn2021probabilistic} the mathematical formulation of these state transitions with new data $\mathbf{Z}_t$ and decisions can be expressed as

\begin{equation}
\label{eq:pdt_full}
\begin{split}
d_t  \propto &\sum_{\textbf{X}_{t-1}}\sum_{\textbf{Q}_{t}} 
    \underbrace{p \left( \textbf{x}_{t-1} | \textbf{z}_{0:t-1}, \textbf{u}_{0:t-2} \right)}_{d_{t-1}} \times\\
    & \underbrace{p\left(\textbf{x}_t | \textbf{x}_{t-1}, \textbf{u}_{t-1} \right)}_{\phi^{transition}} \times \underbrace{p\left( \textbf{z}_{t} | \textbf{x}_{t},\textbf{q}_{t},\textbf{u}_{t-1} \right)}_{\phi^{data}} \times \underbrace{p\left( \textbf{q}_{t}|\textbf{x}_{t} \right)}_{\phi^{pred}} \\
     = &\sum_{\textbf{X}_{t-1}}\sum_{\textbf{Q}_{t}} d_{t-1} \phi^{transition} \phi^{data} \phi^{pred}.
\end{split}
\end{equation}

The current state is dependent on the previous state $d_{t-1}$, which encapsulates the cumulative history up to $t-1$. $\phi^{transition}$ describes the transition from $\mathbf{X}_{t-1}$ to $\mathbf{X}_{t}$ affected by the decisions $\mathbf{u}_{t-1}$. $\phi^{pred}$ describes the dependency of the quantities of interest $\mathbf{Q}_{t}$ on the state $\mathbf{X}_t$, and $\phi^{data}$ represents the likelihood of the measurement data $Z_t$. In \Cref{eq:pdt_full} random variables are assumed to be discrete, but the formulation is general. To include continuous random variables, the summations in the equation must be replaced with corresponding integrals.

\begin{figure*}[h!]
\centering
\includegraphics[width=\textwidth]{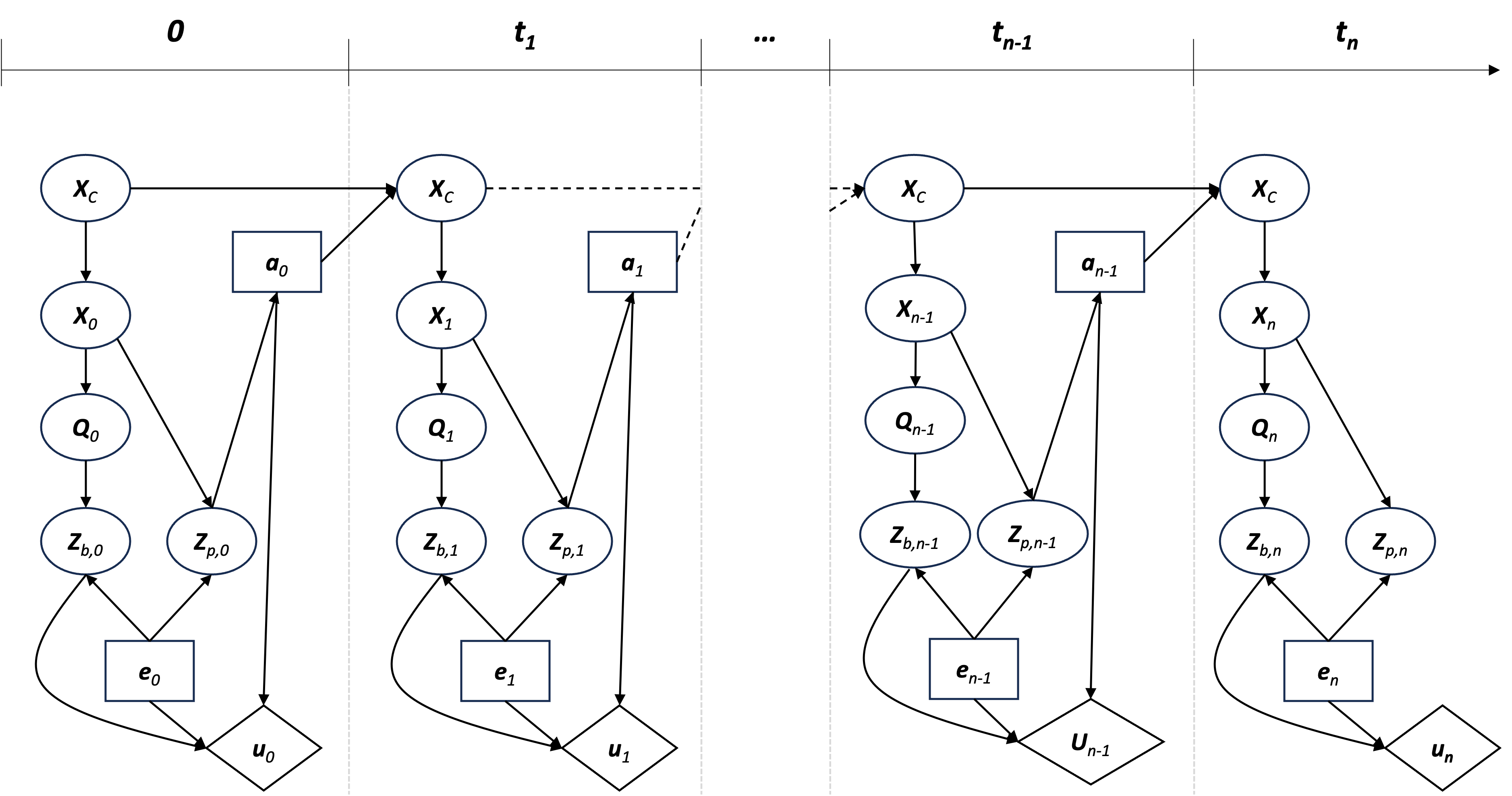}
\caption{A more explicit representation of the PDT model illustrated in \Cref{fig:dt_if_diagram}, which highlights the different types of decisions, data and random variables explicitly.}
\label{fig:dt_if_diagram_extended}
\end{figure*}

We propose an extension of the PDT framework shown in \Cref{fig:dt_if_diagram} to facilitate practical implementation of Bayesian updating in engineering applications. This extension, shown in Figure \ref{fig:dt_if_diagram_extended} is fully compatible with the original formulation. The key modification is that decisions related to information collection, $\mathbf{e}_t$, are assumed to result in measurement data $\mathbf{Z}_t$ within the same time step, rather than at the subsequent time step $\mathbf{Z}_{t+1}$.

We further propose to explicitly distinguish between time-invariant random variables $\mathbf{X}_c$ and time-dependent random variables $\mathbf{X}_{v,t}$. The system state at time $t$ is thus defined as $\textbf{X}_t=\left[ \textbf{X}_{v,t},\textbf{X}_c \right]$. Time-invariant variables are constant over time and represent parameters that can be progressively learned through data assimilation (e.g., soil or material properties). In contrast, time-dependent variables change over time and are here assumed to be independent between time steps. These variables typically represent external influences, such as water levels, traffic loads, or environmental actions.

The extended influence diagram of \Cref{fig:dt_if_diagram_extended} also explicitly distinguishes between property data $\mathbf{Z}_{p,t}$ and behavior data $\mathbf{Z}_{b,t}$. Property data provide direct information on individual random variables or a small subset of the state vector and can be used to update the corresponding marginal distributions by Bayesian analysis without requiring Bayesian inversion. Behavior data, in contrast, are observations of the system performance and are linked to the state through the predicted quantities of interest $\mathbf{Q}_t$. Updating with behavior data, therefore, requires Bayesian inversion using a physics-based or surrogate prediction model. 

Finally, decisions on information collection $\mathbf{e}_t$ and controlling actions $\mathbf{a}_t$ influence the reward $\mathbf{R}_t$. This explicit representation clarifies the role of multi-source data and decision-making within the PDT framework and provides a structured basis for implementing sequential Bayesian updating in probabilistic digital twins. In this work, the focus is on the Bayesian updating process. We do not address decision optimization, which is left for future work.

\section{Updating within the PDT}
\label{sec:pdt_updating}

\subsection{Introduction}
\label{subsec:updating_intro}

For a fixed time step $t$ and given decisions $\mathbf{u}_{0:t}$ and state transitions, the general PDT formulation in \Cref{eq:pdt_full} reduces to the inference of the system state conditional on the available observation data $d_t = p\left( \mathbf{x}_t \mid  \mathbf{z}_{0:t} \right)$, where we drop the dependence on $\mathbf{u}_{0:t}$ for readability. In the following, we switch from the notation used for discrete random variables to that for continuous random variables, which are common in most civil engineering and geotechnical applications. Accordingly, probability mass functions $p(\cdot)$ are replaced by probability density functions (PDF) $f(\cdot)$.

The PDT includes data via Bayes' rule in Eq. \ref{eq:pdt_full}. Specifically, $d_{t-1}$, $\phi^{transition}$ and $\phi^{pred}$ jointly define the prior. The term $\phi^{data}=f\left( \textbf{z}_{t} | \textbf{x}_{t},\textbf{q}_{t} \right)$ represents the likelihood, which corresponds to the probability of the observed data given a set of parameters. The non-trivial part of applying Bayes' rule is the proportionality constant in  Eq. \ref{eq:pdt_full}, which requires advanced methods depending on the type of data. At time $t$, these can be property data $\mathbf{Z}_{p,t}$, behavior data $\mathbf{Z}_{b,t}$, or both, such that $\mathbf{Z}_t=[\mathbf{Z}_{p,t};\mathbf{Z}_{b,t}]$.

For property data $\mathbf{Z}_{p,t}$, each measurement typically corresponds to one or a few random variables in $\mathbf{X}_{t}$. 
The likelihood for property data is given as

\begin{equation}
\label{eq:prop_likelihood_total}
    L_{p,t}(\mathbf{x}_t) = \prod_{k=1}^{n_{prop}} L_{p,t,k}({\mathbf{x}}_{t}),
\end{equation}

where $n_{prop}$ is the number of property data and $L_{p,t,k}$ is the likelihood describing the $k$th observation. Because each individual $L_{p,t,k}$ is related only to one or a few random variables in $\mathbf{X}_{t}$, the proportionality constant in Bayes' rule can be obtained cheaply using methods such as numerical integration (see \Cref{subsec:methods_property_updating}). 

By contrast, behavior data $\mathbf{Z}_{b,t}$ capture system responses that are compared against model predictions. The prediction is obtained as a result of a model evaluation $Q_{t,j}=m_j(\mathbf{X}_{t})$, where $m_j$ is the predictive function e.g., an output of a FEM. A common likelihood function describing a single observation is 

\begin{equation}
\label{eq:beh_likelihood_individual}
    L_{b,t,j}(\mathbf{x}_{t}) = f_{\epsilon_j}\left( z_{b,t,j}-m_j(\mathbf{x}_{t})\right),
 \end{equation}
where $z_{b,t,j}$ is the $j$th observation and $f_{\epsilon_j}$ is the PDF of the measurement error.

The joint likelihood of all behavior data at time $t$ is

\begin{equation}
\label{eq:beh_likelihood_total}
    L_{b,t}(\mathbf{x}_t) = \prod_{j=1}^{n_{beh}} L_{b,t,j}(\mathbf{x}_t).
\end{equation}
where $n_{beh}$ is the number of behavior data.

Updating in this context is challenging, as we generally deal with high-dimensional spaces in which many RVs influence the prediction of the quantity of interest. Here we use sampling-based methods to evaluate the posterior, as described in \Cref{subsec:methods_behavior_updating}.

\subsection{Two-step Bayesian updating within the PDT}
\label{subsec:pdt_cont_updating}

To take advantage of the fact that different data types affect distinct subsets of the digital state $\mathbf{X}_t$ and enable different numerical treatments, we propose a two-step updating scheme that combines multiple methods in one framework. This scheme deviates from the sequential updating scheme in Eq. \ref{eq:pdt_full} because the Bayesian updating is not done in a sequential manner. The main reason for this deviation from the original PDT formulation is that the numerical solutions can deteriorate when using a sequential approach. To ensure computational efficiency, instead a surrogate model is learned sequentially, which is described in \Cref{sec:seq_gpr}.

In the first step of the proposed scheme, the distribution of $\mathbf{X}_{0:t}$ is updated using all available property data. These updates typically occur in low-dimensional spaces because most measurements relate to only a small number of random variables. At each time step $t$, the prior of $\mathbf{X}$ is updated with the full history of property data $\mathbf{Z}_{p,0:t}$. Thus, we obtain the posterior for all property data as

\begin{equation}
\label{eq:property_updating}
     f(\mathbf{x}_{0:t}|\mathbf{z}_{p,0:t}) \propto f(\mathbf{x}_{0:t}) \prod_{i=0}^t L_{p,i}(\mathbf{x}_t).
\end{equation}

Because in most cases, property data is only informative on a subset $\mathbf{X}_{p,0:t}$ of the random variables, we rewrite this to

\begin{equation}
\label{eq:property_updating_subset}
     f(\mathbf{x}_{0:t}|\mathbf{z}_{p,0:t}) \propto f(\mathbf{x}_{p,0:t})f(\mathbf{x}_{\bar{p},0:t}|\mathbf{x}_{p,0:t}) \prod_{i=0}^t L_{p,i}(\mathbf{x}_{p,t}).
\end{equation}
in which $f(\mathbf{x}_{p,0:t})$ is the PDF of the random variables in $\mathbf{X}$ that are related to the property data and $f(\mathbf{x}_{\bar{p},0:t}|\mathbf{x}_{p,0:t})$ is the conditional PDF of the remaining random variables in $\mathbf{X}$.

In the second step, $f\left( \mathbf{x}_t|\mathbf{z}_{p,0:t} \right)$ is updated using the full history of behavioral data $\mathbf{z}^{beh}_{0:t}$ up to time $t$ by

\begin{equation}
\label{eq:behavior_updating_batch}
f(\mathbf{x}_t|\mathbf{z}_{p,0:t},\mathbf{z}_{b,0:t}) \propto f(\mathbf{x}_{0:t}|\mathbf{z}_{p,0:t}) \prod_{i=0}^{t} L_{b,i}(\mathbf{x}_i).
\end{equation}

In \Cref{eq:behavior_updating_batch}, the full history of behavioral likelihoods is evaluated at each time step. In the following, this approach is referred to as \emph{batch updating}. 

Alternatively, we also consider sequential inclusion of behavioral data, in line with the original PDT formulation. In this case, the posterior after each behavioral update is used as the prior for the next one, while the full history of property data $\mathbf{z}_{p,0:t}$ is retained. This gives the updating scheme

\begin{equation}
\label{eq:behavior_updating_seq}
\begin{aligned}
f(\mathbf{x}_t|\mathbf{z}_{p,0:t},\mathbf{z}_{b,0}) 
&\propto f(\mathbf{X}_0|\mathbf{z}_{p,0:t}) L_{b,0}(\mathbf{x}_0), \\
f(\mathbf{x}_t|\mathbf{z}_{p,0:t},\mathbf{z}_{b,0:1})
&\propto f(\mathbf{x}_1|\mathbf{z}_{p,0:t},\mathbf{z}_{b,0}) L_{b,1}(\mathbf{x}_1), \\ &\vdots \\
f(\mathbf{x}_t|\mathbf{z}_{p,0:t},\mathbf{z}_{b,0:t})
&\propto f(\mathbf{x}_t|\mathbf{z}_{p,0:t},\mathbf{z}_{b,0:t-1}) L_{b,t}(\mathbf{x}_t).
\end{aligned}
\end{equation}

We refer to this approach as \emph{sequential updating}. Compared with batch updating, this formulation can be advantageous when property data is available only infrequently, e.g., at the beginning of the service life. In these cases, the approach can reduce computational cost. However, in sampling-based inference, approximation errors may accumulate over sequential updates. At each time step, the posterior distribution is approximated by a finite set of samples. The resulting error may propagate to subsequent updates and compound. For this reason, both batch and sequential updating are investigated in this work.

\subsection{Updating with Property Data}
\label{subsec:methods_property_updating}

Property data mostly relate to a single random variable or a small subset of the system state $\mathbf{X}_t$. As a result, Bayesian updating with property data often takes place in a low-dimensional parameter space. The proportionality constant in \Cref{eq:property_updating_subset} is

\begin{equation}
\label{eq:property_updating_constant}
     \text{constant} = \left[ \int_{\mathbf{X}_p} f(\mathbf{x}_{p,0:t}) \prod_{i=0}^t L_{p,i}(\mathbf{x}_{p,t}) \text{d}\mathbf{x}_{p} \right]^{-1} .
\end{equation}

If $\mathbf{X}_p$ is low-dimensional, the posterior distribution can be obtained using standard numerical integration methods, such as the trapezoidal rule, Simpson's rule, or Gaussian quadrature. 
An advantage of this approach is that the posterior distribution remains available in analytical form.
 
However, property data may also relate to spatially distributed properties. A common example in geotechnical engineering is cone penetration test (CPT) data, which result from measurements of soil properties with high spatial resolution. 

In such cases, methods based on Gaussian process regression (GPR) are well-suited because they naturally support Bayesian updating and can efficiently incorporate large datasets while accounting for spatial correlation. Many examples of GP-based updating in geotechnical applications can be found in the literature \citep[e.g.,][]{Li2016,yoshida2021estimation,zinas_multivariate_2025}.

\subsection{Methods for Updating with Behavior Data}
\label{subsec:methods_behavior_updating}

Solving either \Cref{eq:behavior_updating_seq} or \Cref{eq:behavior_updating_batch} is challenging, as evaluating the likelihood is expensive and requires running the underlying prediction model. This constitutes a Bayesian inversion problem for which analytical solutions are unavailable and standard numerical integration is generally infeasible due to the curse of dimensionality. Consequently, sampling-based methods have been developed to enable direct sampling from the posterior distribution. Here, we employ  the BUS method (Bayesian Updating with Structural Reliability Methods), proposed in \citep{straub2011reliability,straub2015bayesian}, which reformulates the Bayesian inversion as a reliability problem.

In the  BUS framework, the behaviour data $\mathbf{z}_{b,0:t}$ is described by a limit state function 
\begin{equation}
\label{eq:BUS_LSF}
\begin{aligned}
h(\mathbf{x}_{0:t},p)&=\ln p-\ln c -\ln \left[\prod_{i=0}^{t} L_{b,i}(\mathbf{x}_i)\right] \\ &= \ln p-\ln c -\sum_{i=0}^{t} \ln L_{b,i}(\mathbf{x}_i),
\end{aligned}
\end{equation}
in which $P\sim U(0,1)$ is an auxiliary uniform random variable and $c$ is a scaling constant ensuring $c \cdot \prod_{i=0}^{t} L_{b,i}(\mathbf{x}_i) \leq 1$ for all $\mathbf{x}$. 
Conditioning on $\mathbf{z}_{b,0:t}$ is achieved by evaluating samples conditional on $h(\mathbf{x}_{0:t},p)\le 0$.
The BUS method is particularly efficient in reliability problems.For details on the implementation, see \citet{straub2015bayesian,diazdelao_bayesian_2017,betz2018bayesian}. 

The computational cost of BUS is primarily determined by the number of likelihood evaluations. For behavioral data, where measurements are related to prediction model outputs, evaluating the likelihood can be expensive. This motivates the use of surrogate models and active learning, as discussed in the following section.

\section{Sequential Surrogate Model for Bayesian Inversion}
\label{sec:seq_gpr}

\subsection{Introduction}
\label{subsec:seq_gpr_motivation}

Surrogate modeling provides a computationally efficient method to approximate the response of a physics-based model with only a few expensive model evaluations. Common techniques include polynomial chaos expansion, support vector regression, neural networks, and GPR. Active learning can further improve efficiency by focusing model evaluations on regions of interest. This is of particular relevance in Bayesian updating and reliability analysis, where active learning surrogates are popular \cite{echard_ak-mcs_2011,moustapha_active_2022}.
The integration of active learning with BUS has been studied in the literature \citep[][]{wang_highly_2020,song_buak-ais_2022,rossat_bayesian_2022,jiang_adaptive_2023,prentzas_active_2025}. Existing approaches mainly differ in their learning functions and stopping criteria, but typically approximate the limit state function directly.

A challenge in the digital twin context is that both the system state $\mathbf{X}_t$ and the measurement data $\mathbf{Z}_t$ evolve over time. The problem is therefore related to time-variant reliability analysis \citep{mori1993reliability,rackwitz1998computational,straub2020reliability}, and involves shifting regions of interest for the prediction model as parameters are updated, components deteriorate, or loading conditions change. 

\begin{figure*}[t]
\centering
\includegraphics[width=0.9\textwidth]{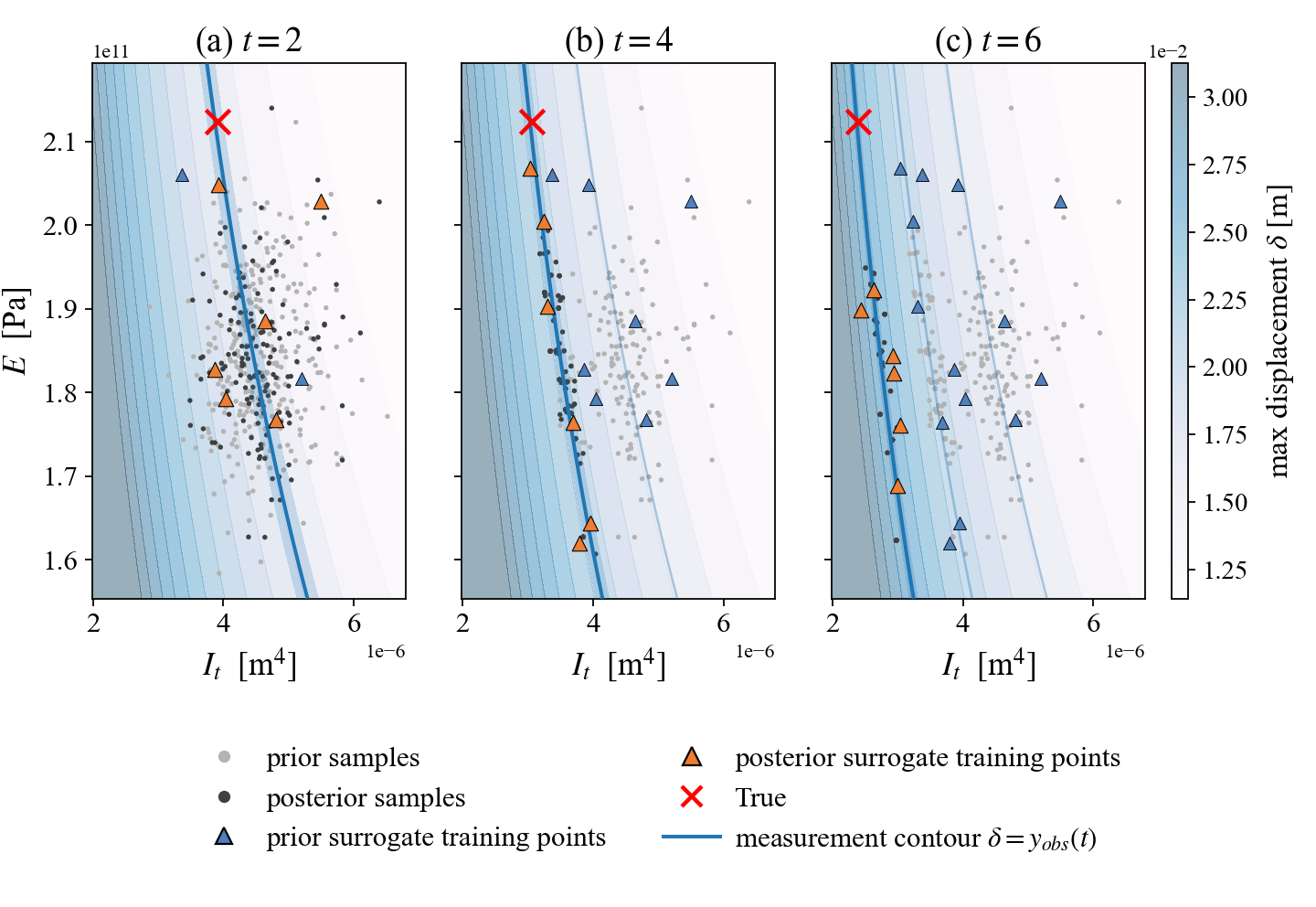}
\caption{Evolution of the true state over time and additional training points used to refine the surrogate model.}
\label{fig:seq_sur_workflow_true_state}
\end{figure*}

To illustrate this, consider a cantilever beam subjected to a distributed load $V$, with elastic modulus $E$ and a moment of inertia $I_t=I_{start}*\lambda^t$ that depends on time $t$ through the deterioration parameter $\lambda$. $V$ is assumed to be deterministic, so that the random variables of the problem are $\mathbf{X}_t=[I_t,E]$. We assume that displacement measurements are obtained at $t=[2,4,6]$ years, and Bayesian inversion can be used to update the distributions of $\mathbf{X}_t$. A surrogate model with $\mathbf{X}_t$ as input is used to predict the maximum displacement at the free end of the beam. In this setting, the same surrogate model is reused at different time steps, while the relevant region of the input space changes as the moment of inertia evolves according to $I_t$. This is illustrated in \Cref{fig:seq_sur_workflow_true_state}. As shown there, a surrogate trained for the region relevant at $t=2$ may not be sufficiently accurate in the region relevant at $t=4$.

To avoid constructing separate time-invariant surrogate models for each time step, we propose a sequential surrogate modeling approach for Bayesian inversion within a probabilistic digital twin framework. In contrast to approaches that approximate the limit state functions directly, the output quantity of the surrogate is a vector of physical quantities of interest $Q_t$. This enables the same surrogate to be used for Bayesian updating and reliability analysis at different times $t$. The approach combines GPR, active learning strategies, and sampling-based Bayesian updating, allowing the surrogate model to evolve alongside the digital twin as new data is obtained.

\subsection{Sequential surrogate}
\label{subsec:workflow}

In active learning, the training set of a surrogate model is enriched sequentially until a prescribed accuracy is reached. This requires a \emph{learning function}, which selects the next training point, and a \emph{stopping criterion}, which indicates convergence. An overview of learning functions and stopping criteria in active-learning reliability is provided by \citet{moustapha_active_2022}.

A widely used learning function in active-learning reliability methods with GPR is the $U$-function \citep{echard_ak-mcs_2011}. For a limit state function $h$, the $U$-function for any candidate point $\mathbf{x}$ is
\begin{equation}
\label{eq:u_learning}
    U(\mathbf{x}) = \frac{\left\lvert\mu_{\hat{h}}(\mathbf{x})\right\rvert}{\sigma_{\hat{h}}(\mathbf{x})},
\end{equation}
where $\mu_{\hat{h}}(\mathbf{x})$ and $\sigma_{\hat{h}}(\mathbf{x})$ are the mean and standard deviation of the surrogate model that approximates the limit state function $h(\mathbf{x})$. Small values of $U$ indicate candidate points close to the limit state surface or in regions of high surrogate uncertainty.

In the present work, the surrogate is not used to approximate the limit state function $h$ directly. Instead, it approximates one or more outputs of a physics-based response model, $m$. A surrogate-induced limit state function $\hat{h}$ is then constructed from the surrogate response model $\hat{m}$. Consequently, the predictive mean and standard deviation of $\hat{h}$, which are required to evaluate the $U$-function in \Cref{eq:u_learning}, must be obtained from the predictive distribution of $\hat{m}$. For nonlinear transformations from $m$ to $h$, these moments are approximated using a first-order Taylor expansion. The performance of the active learning approach under this approximation is assessed in the numerical examples.

For Bayesian updating, we use the BUS formulation presented in \Cref{eq:BUS_LSF}. The surrogate approximation of the likelihood $\prod_{i=0}^{t} \hat{L}_{b,i}(\mathbf{x}_i)$ is obtained by replacing the outputs of the physics model $m_j$ with the corresponding surrogate model output $\hat{m}_j$ in Equations \ref{eq:beh_likelihood_individual} and \ref{eq:beh_likelihood_total}. The resulting surrogate-induced BUS limit state function is written in logarithmic form by inserting this likelihood into \Cref{eq:BUS_LSF}:
\begin{equation}
\label{eq:bus_lsf_surrogate}
    \hat{h}(\mathbf{x},p)
    =
    \ln p - \ln c -  \sum_{i=0}^{t} \ln \hat{L}_{b,i}(\mathbf{x}_i) ,
\end{equation}

To obtain samples conditional on $\hat{h}(\mathbf{x},p)\le 0$, we use Subset Simulation, wherein the posterior domain is reached through a sequence of intermediate domains
\begin{equation}
\label{eq:subset_domains}
    \Omega_k =
    \left\{
    (\mathbf{x},p): \hat{h}(\mathbf{x},p) \leq b_k
    \right\},
\end{equation}
with $b_k$ chosen such that a specified fraction $p_0$ of the samples satisfies this condition \citep{au2001estimation,au2007application,papaioannou2015mcmc,betz2018bayesian}. The final target domain corresponds to $b_k=0$. To apply the $U$-function at subset level $k$, we define the shifted surrogate limit state function
\begin{equation}
\label{eq:shifted_h}
\begin{split}
    \hat{h}_k(\mathbf{x},p)
    &=
    \hat{h}(\mathbf{x},p)-b_k \\
    =& \ln p - \ln c - \sum_{i=0}^{t} \ln \hat{L}_{b,i}(\mathbf{x}_i)-b_k.
\end{split}
\end{equation}

The corresponding predictive mean and standard deviation are

\begin{subequations}\label{eq:h_mean_std}
\begin{align}
\mu_{\hat{h}_k}(\mathbf{x},p)
  &= \ln p - \ln c - \mu_{\ln \hat{L}_{b,i}}(\mathbf{x}) - b_k,
  \label{eq:h_mean_std_a}\\
\sigma_{\hat{h}_k}(\mathbf{x},p)
  &= \sigma_{\ln \hat{L}_{b,i}}(\mathbf{x}).
  \label{eq:h_mean_std_b}
\end{align}
\end{subequations}

where $\mu_{\ln \hat{L}_{b,i}}(\mathbf{x})$ and $\sigma_{\ln \hat{L}_{b,i}}(\mathbf{x})$ are the mean and standard deviation of $\ln \hat{L}_{b,i}(\mathbf{x})=\sum_{i=0}^{t} \ln \hat{L}_{b,i}(\mathbf{x}_i)$.

The log-likelihood in function of $\hat m(\mathbf{x})$ is approximated by a first-order Taylor expansion about $\mu_{\hat m}(\mathbf{x})$:
\begin{equation}
\begin{split}
\ln \hat L_{b,i}(\mathbf{x})&=\ln L_{b,i}\!\left(z\mid\hat m(\mathbf{x})\right)\\&\approx\ln L_{b,i}\!\left(z\mid\mu_{\hat m}(\mathbf{x})\right)+\left.
\frac{\partial\ln L_{b,i}\left(z\mid m\right)}{\partial m}\right|_{m=\mu_{\hat m}(\mathbf{x})}\left[\hat m(\mathbf{x})-\mu_{\hat m}(\mathbf{x})\right].
\end{split}
\label{eq:log_likelihood_taylor}
\end{equation}

The resulting approximate mean and standard deviation of $\ln \hat L_{b,i}(\mathbf{x})$ are
\begin{subequations}
\label{eq:log_likelihood_moments}
\begin{align}
\mu_{\ln \hat L_{b,i}}(\mathbf{x})&\approx\ln L_{b,i}\!\left(z\mid\mu_{\hat m}(\mathbf{x})\right),\label{eq:log_likelihood_moments_a}\\\sigma_{\ln \hat L_{b,i}}(\mathbf{x})&\approx\left|\left.\frac{\partial\ln L_{b,i}\left(z\mid m\right)}{\partial m}\right|_{m=\mu_{\hat m}(\mathbf{x})}\right|\sigma_{\hat m}(\mathbf{x}).
\label{eq:log_likelihood_moments_b}
\end{align}
\end{subequations}

The active-learning function in \Cref{eq:u_learning} is evaluated using $\hat{h}_k$ at the current subset level. To define a convergence criterion, the fraction of candidate samples in $\Omega_{k-1}$ satisfying the current subset condition is used to estimate the conditional probability 

\begin{equation}
\label{eq:conditional_bus_probability}
P_k=P\left(\hat{h}_k(\mathbf{x},p)\leq0\mid(\mathbf{x},p)\in\Omega_{k-1}\right)=
P\left(\Omega_k\mid\Omega_{k-1}\right).
\end{equation}

For a fixed current subset boundary, the active-learning procedure is considered converged when the relative change in the estimate $\hat{P}_k^{(j)}$ between two successive active-learning iterations $j$ falls below a prescribed threshold $\epsilon_P$

\begin{equation}
\label{eq:al_stopping_criterion}
\frac{\left|\hat{P}_k^{(j)}-\hat{P}_k^{(j-1)}\right|
}{\hat{P}_k^{(j)}}\leq \epsilon_P.
\end{equation}

This criterion is analogous to the reliability-index stability criterion discussed by \citet{moustapha_active_2022}. If the criterion is not satisfied, the next training point is selected as
\begin{equation}
\label{eq:u_argmin}
    (\mathbf{x}^{\ast},p^{\ast})
    =\arg\min_{(\mathbf{x},p)\in\mathcal{C}} U_k(\mathbf{x},p)    =    \arg\min_{(\mathbf{x},p)\in\mathcal{C}} \frac{\ln p - \ln c - \mu_{\ln \hat{L}}(\mathbf{x})  - b_k}{\sigma_{\ln \hat{L}}(\mathbf{x})}.
\end{equation}

The selected model evaluation is added to the training set, and the surrogate is refitted. The procedure is repeated until \Cref{eq:al_stopping_criterion} is satisfied or the maximum number of active-learning iterations $n_{\mathrm{max,it}}$ is reached. 

The resulting active-learning refinement procedure is summarized in \Cref{alg:al_only}. Although the $U$-function and the stopping criterion in \Cref{eq:al_stopping_criterion} are used in this work, the workflow is not restricted to these choices. Other learning functions and convergence criteria can be incorporated in the same framework. The formulation is also not limited to Subset Simulation. For sampling methods that do not use intermediate thresholds, the same learning function can be applied directly to the final BUS boundary by setting $b_k=0$. 

\begin{algorithm}[h!]
\caption{Sequential surrogate refinement for Bayesian inversion at subset level $k$.}
\label{alg:al_only}
\SetKwInOut{Input}{input}
\SetKwInOut{Output}{output}
\SetKwFunction{Converged}{Converged}
\SetKwData{True}{true}
\SetAlgoLined

\Input{Starting surrogate prediction model $\hat{m}_{k}^{(0)}$;
physics-based prediction model $m$;
current training set $\mathcal{T}$;
candidate samples $(\mathbf{x},p)\in\mathcal{C}$ from $\Omega_{k-1}$, $N=\lvert\mathcal{C}\rvert$;
subset simulation parameter $p_{0}$;
BUS scaling constant $c^{(0)}$;
convergence tolerance $\epsilon_{P}$;
maximum number of iterations $n_{\mathrm{max,it}}$.}

\Output{Refined surrogate model $\hat{m}_{k}^{(j)}$.}

\For{$j \leftarrow 1$ \KwTo $n_{\mathrm{max,it}}$}{

  \ForEach{$(\mathbf{x},p)\in\mathcal{C}$}{
    %Obtain $\mu_{\ln \hat{L}}^{(j)}(\mathbf{x})$ and $\sigma_{\ln \hat{L}}^{(j)}(\mathbf{x})$ from the current surrogate $\hat{m}_{k}^{(j)}$\;
    $\mu_{\ln \hat L}^{(j)}(\mathbf{x}) \approx\ln L\!\left(z\mid\mu_{\hat{m}_{k}^{(j)}}(\mathbf{x})\right)$\;
    $\sigma_{\ln \hat L}^{(j)}(\mathbf{x}) \approx\left|\left.\frac{\partial\ln L\left(z\mid m \right)}{\partial m}\right|_{m=\mu_{\hat{m}_{k}^{(j)}}(\mathbf{x})}\right|\sigma_{\hat{m}_{k}^{(j)}}(\mathbf{x})$\;
  }

  $\ln c^{(j)} \leftarrow
   \min\!\Bigl(\ln c^{(j-1)},\;
   -\max_{(\mathbf{x},p)\in\mathcal{C}} \mu_{\ln \hat{L}}^{(j)}(\mathbf{x})\Bigr)$\;

  \ForEach{$(\mathbf{x},p)\in\mathcal{C}$}{
    $\mu_{\hat{h}_k}^{(j)}(\mathbf{x},p) \leftarrow
     \ln p - \ln c^{(j)} - \mu_{\ln \hat{L}}^{(j)}(\mathbf{x})$\;
    $\sigma_{\hat{h}_k}^{(j)}(\mathbf{x},p) \leftarrow
     \sigma_{\ln \hat{L}}^{(j)}(\mathbf{x})$\;
  }

   $\bigl(v_{(1)}^{(j)},\ldots,v_{(N)}^{(j)}\bigr)
 \leftarrow \operatorname{sort}_{\uparrow}
 \left\{\mu_{\hat h_k}^{(j)}(\mathbf{x},p):
        (\mathbf{x},p)\in\mathcal{C}\right\}$\;
     $b_k^{(j)} \leftarrow
 \max\!\left(0,\,
 \frac{v_{(Np_0)}^{(j)}+v_{(Np_0+1)}^{(j)}}{2}\right)$\;

  $\hat{P}_{k}^{(j)} \leftarrow
   \dfrac{1}{N}\sum_{(\mathbf{x},p)\in\mathcal{C}}
   \mathbb{I}\!\left[\mu_{\hat{h}_k}^{(j)}(\mathbf{x},p) - b_{k}^{(j)} \leq 0\right]$\;

    \If{$j \geq 1$ \textbf{and}
    $\dfrac{\bigl|\hat{P}_k^{(j)} - \hat{P}_k^{(j-1)}\bigr|}{\hat{P}_f^{(j)}}
    \leq \epsilon_{P}$}{
    \Return $\hat{m}_{k}^{(j)}$\;
    }

  \ForEach{$(\mathbf{x},p)\in\mathcal{C}$}{
    $U_{k}^{(j)}(\mathbf{x},p) \leftarrow
     \dfrac{\bigl\lvert \mu_{\hat{h}_k}^{(j)}(\mathbf{x},p) - b_{k}^{(j)} \bigr\rvert}
           {\sigma_{\hat{h}_k}^{(j)}(\mathbf{x},p)}$\;
  }

  $(\mathbf{x}^{\star},p^{\star}) \leftarrow
   \arg\min_{(\mathbf{x},p)\in\mathcal{C}} U_{k}^{(j)}(\mathbf{x},p)$\;
  $y^{\star} \leftarrow m(\mathbf{x}^{\star})$\;
  $\mathcal{T} \leftarrow \mathcal{T} \cup \{(\mathbf{x}^{\star},y^{\star})\}$\;
  Retrain the surrogate on $\mathcal{T}$ to obtain $\hat{m}_{k}^{(j+1)}$\;

  \If{$j = n_{\mathrm{max,it}}-1$}{
    \Return $\hat{m}_{k}^{(j+1)}$\;
  }
}
\end{algorithm}
\clearpage

\section{Numerical Investigations: Cantilever Beam}
\label{sec:example_cantilever}

\subsection{Setup}
\label{subsec:beam_intro}

\begin{figure*}[h]
\centering
\begin{subfigure}[]
    \centering
    \includegraphics[width=0.3\textwidth]{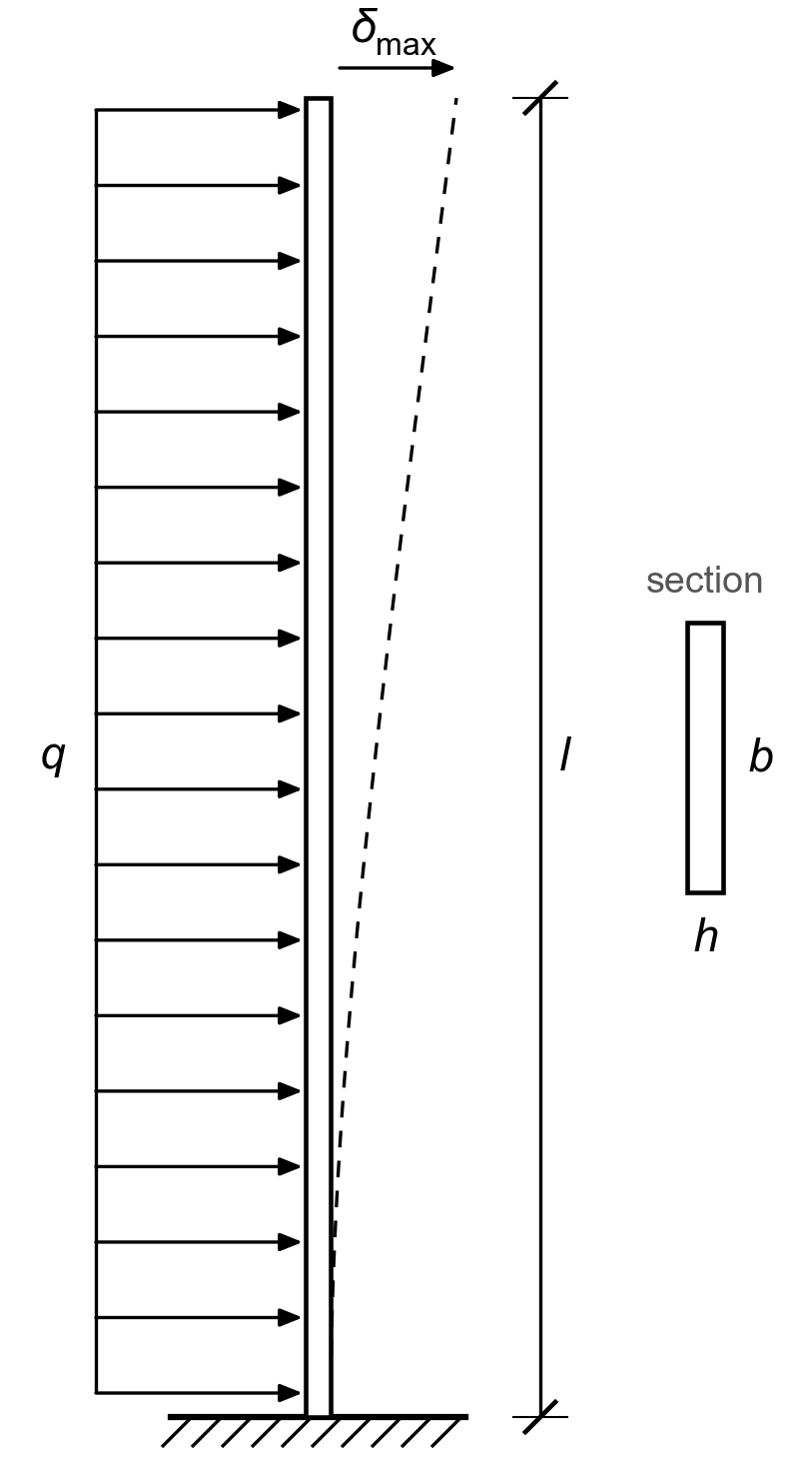}
    % \caption{Simply Supported Beam }
    \label{fig:cantilever_beam_dimensions}
\end{subfigure}
% \hfill
\begin{subfigure}[]
    \centering
    \includegraphics[width=0.4\textwidth]{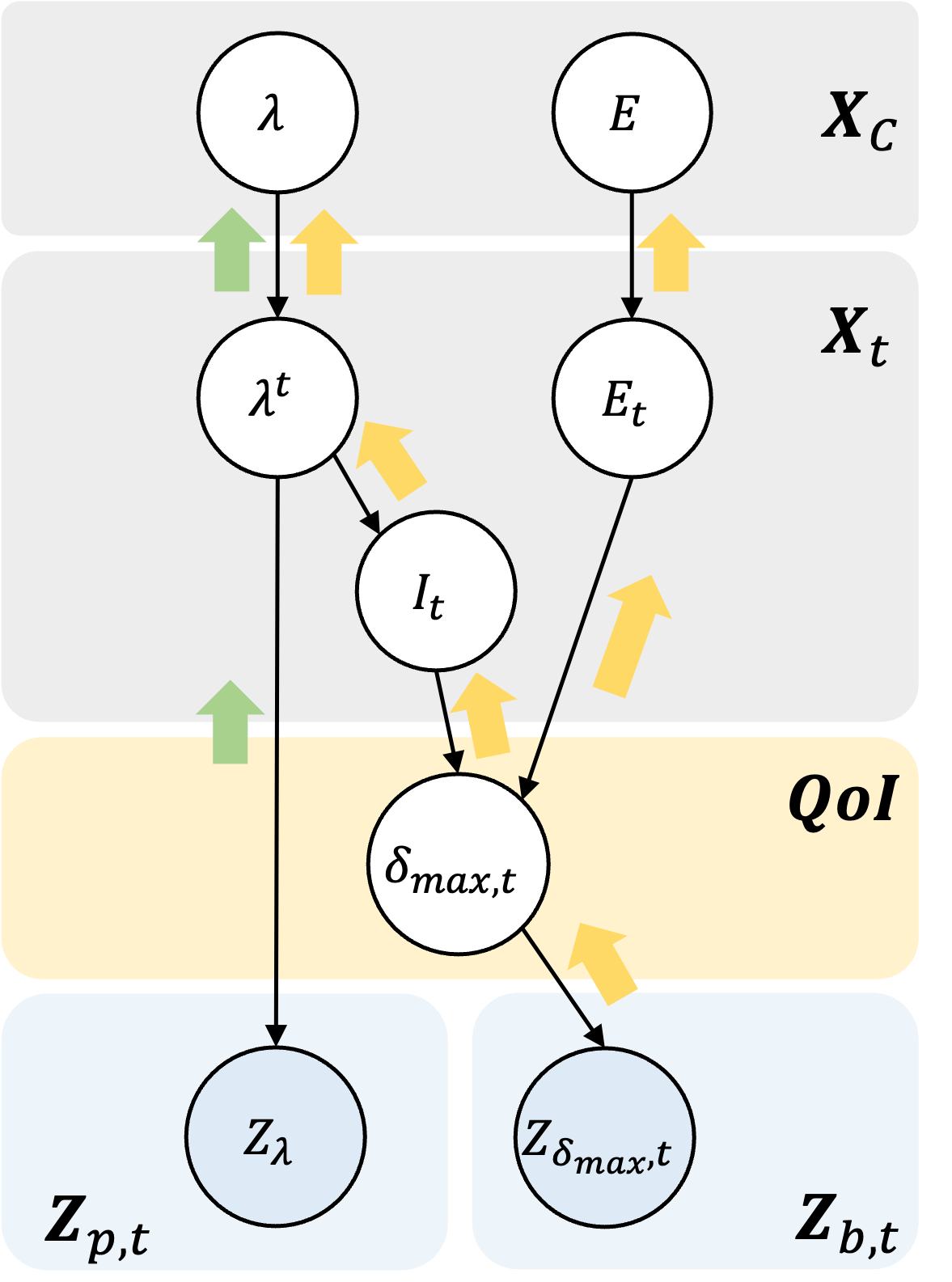}
    % \caption{Bayesian network of the cantilever beam}
    \label{fig:cantilever_beam_information_flow}
\end{subfigure}
\caption{(a) Cantilever beam  (b) Bayesian network of the cantilever beam}
\label{fig:cantilever_beam}
\end{figure*}

To investigate the proposed method on a basic example, we consider a vertical cantilever beam (see \Cref{fig:cantilever_beam}a). The beam is made of steel, which deteriorates over time. This example includes only a few random variables, and the physical model has analytical solutions. This facilitates the computation of reference solutions and enables extensive numerical evaluations of the proposed method. 

We assume a total service life of $20$ years, with displacement measurements obtained annually and deterioration measurements every $5$ years. The model parameters and their prior distributions are listed in \Cref{tab:beam_parameters}. The setup is described below. The accompanying repository provides the code and configuration files for reproducing the cantilever beam example and exploring alternative settings (see the Code availability statement).

\paragraph{Beam properties}
The beam geometry is defined by its height $L$ and moment of inertia $I$, which depend on the cross-sectional dimensions (width and length). The material behavior is characterized by the Young’s modulus $E$ of steel.

\paragraph{Deterioration}
The beam deteriorates, which reduces the effective cross-sectional area and thus the moment of inertia $I$ (see \Cref{fig:cantilever_beam}b). The moment of inertia thus becomes a function of time and is given as

\begin{equation}
\label{eq:det_rate}
    I(t) = I_0\left(1 - \frac{\lambda}{100}\right)^{t}
\end{equation}

where $\lambda$ is a deterioration parameter modelled by a lognormal distribution and $I_0 = bh^3/12$ is the initial moment of inertia. 

\paragraph{Load Scenario}
    The beam is subjected to a constant line load $V$ applied over its entire length.

\paragraph{Displacement Prediction}
    The displacement at the free end is predicted using classical beam theory as $\delta_{max}=g(\mathbf{X})=Vl^4/8EI$. For the purpose of evaluating the proposed approach, we use a GPR surrogate model with active learning, trained on data generated by this analytical model.

\paragraph{Measurement Data}
The displacement prediction can be improved by collecting more information on the state of the beam.

At each timestep, two types of measurement data can be obtained:
    \begin{enumerate}
    \item Deterioration.\\
    We assume that the reduction in the moment of inertia relative to We assume that the reduction in the moment of inertia relative to $I_0$ can be measured, giving the measured moment of inertia $z_{I,t}$ at time $t$. It is transformed back to $\lambda$ using the inverse of \Cref{eq:det_rate}, yielding $z_{\lambda,t}=100\left[1-(z_{I,t}/I_0)^{1/t}\right]$. For an observed value $z_{\lambda,t}$, the likelihood is

\begin{equation}
    L_{\lambda,t}(\lambda)
    =
    f_{\varepsilon_\lambda}\!\left(z_{\lambda,t}-\lambda\right),
\end{equation}
where $f_{\varepsilon_\lambda}$ is PDF of a normally distributed measurement error with zero mean and standard deviation $\sigma_\lambda$. The standard deviation is fixed at $\sigma_\lambda=0.1$ for all measurement times, corresponding to a measurement uncertainty of roughly $10\%$ of the deterioration rate considered here. Synthetic measurements are generated by adding zero-mean Gaussian noise with a standard deviation equal to $10\%$ of the true reduction $I_0 - I(t)$ to the true moment of inertia $I(t)$ before applying the inverse transformation.

    \item Displacement.\\ 
    The measured displacement at the free end, $Z_{\delta_{\max},t}$, is subject to additive measurement error. For an observed value $z_{\delta_{\max},t}$, the assumed likelihood is
    \begin{equation}
        L_{\delta_{\max},t}(E,\lambda,\mathbf{x})
        =f_{\varepsilon_{\delta_{\max},t}}\!\left(
            z_{\delta_{\max},t}-   g\left(E,\lambda,t,\mathbf{x}\right) \right),
    \end{equation}
    where $f_{\varepsilon_{t,\delta_{\max}}}$ is the PDF of a zero-mean normal error with constant standard deviation $\sigma_{\delta_{\max}}=0.02\,\mathrm{m}$:
    \begin{equation}
    \varepsilon_{\delta_{\max},t}
    \sim
    \mathcal{N}\!\left(0,\sigma_{\delta_{\max}}\right).
    \end{equation}
Synthetic measurements are generated by adding zero-mean Gaussian noise to the true displacement $\delta_{\max}^{\mathrm{true}}(t)$, with a standard deviation equal to $1\%$ of the true displacement at that time. Therefore, the assumed likelihood used in the digital twin differs from the true likelihood, to reflect model inaccuracies that are part of every real-world application.
\end{enumerate}
        
\paragraph{Updating procedure}
    At each time step, Bayesian updating is performed in two stages, following \Cref{subsec:pdt_cont_updating}. First, given a deterioration-rate measurement, $\lambda$ is updated using all available property data by numerical integration. Second, given a displacement measurement, the joint distribution of $(E,\lambda)$ is updated using the aBUS method, following either the batch or the sequential approach.

\begin{table*}[htbp]
      \caption{Cantilever beam parameters and prior distributions.}
      \label{tab:beam_parameters}
      \centering
      \begin{adjustbox}{width=\textwidth}
      \begin{tabular}{l|c|c|c|c|c}
          \hline
          \textbf{Parameter}
          & \textbf{Symbol}
          & \textbf{Unit}
          & \textbf{Distribution}
          & \shortstack{\textbf{Mean/}\\ \textbf{Value}}
          & \shortstack{\textbf{Standard}\\\textbf{deviation}} \\
          \hline
          Beam length
          & $l$
          & $\mathrm{m}$
          & Deterministic
          & $10$
          & -- \\

          Beam width
          & $b$
          & $\mathrm{m}$
          & Deterministic
          & $1.0$
          & -- \\

          Beam height
          & $h$
          & $\mathrm{m}$
          & Deterministic
          & $0.13$
          & -- \\

          Line load
          & $V$
          & $\mathrm{kN}/\mathrm{m}$
          & Deterministic
          & $4$                                  
          & -- \\

          \hline

          Elastic modulus
          & $E$
          & $\mathrm{kN}/\mathrm{m}^2$
          & Normal
          & $17\times10^7$
          & $2.55\times10^7$ \\

          Deterioration parameter
          & $\lambda$
          & $\%/\mathrm{year}$
          & Normal
          & $0.8$
          & $0.16$ \\

          % \hline
  % Observed Deterioration & $Z_{t,DR}$ & Normal & $\mathcal{N}(DR_{t,true},0.1)$ & $[\%/year]^t$\\
  % Observed Displacement & $Z_{t,\delta_{max}}$ & Normal &  $\mathcal{N}(\delta_{max,t,true},0.1)$ & $[m]$\\
          \hline
      \end{tabular}
      \end{adjustbox}
\end{table*}

\subsection{Results}
\label{subsec:beam_results}

\subsubsection{Updating with Property Data}

\begin{figure*}[h]
\centering
\includegraphics[width=\textwidth]{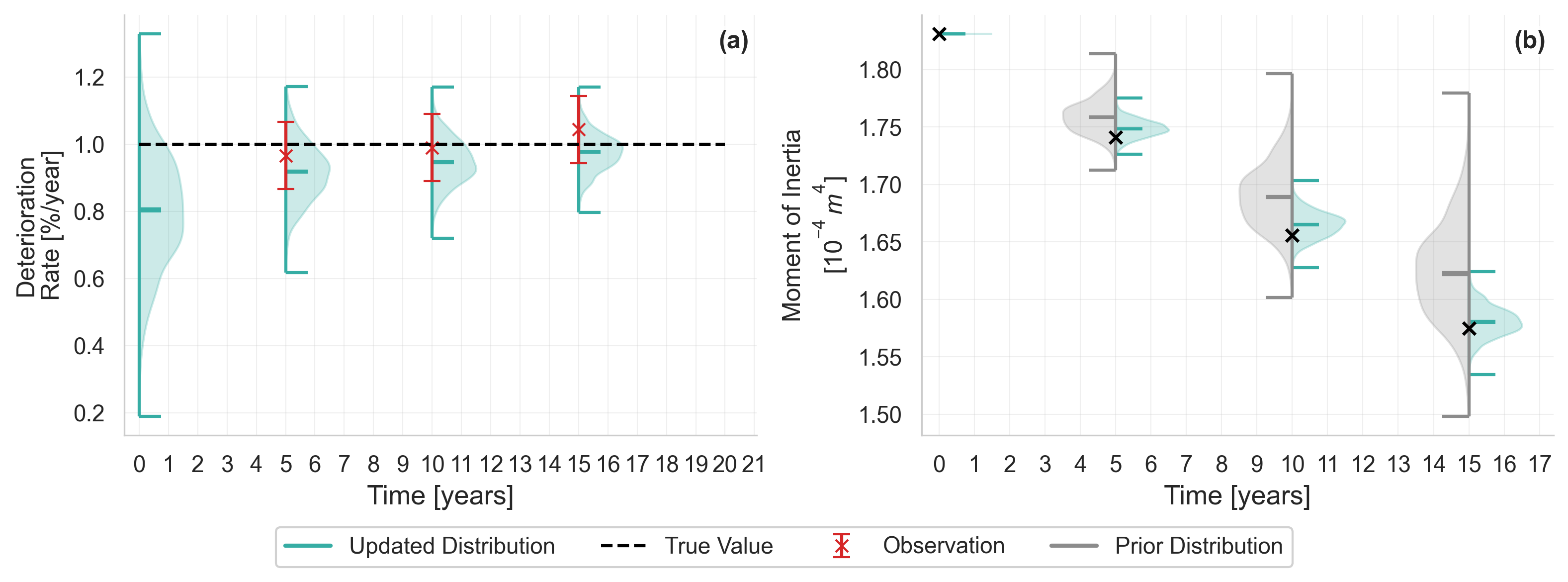}
\caption{Posterior distributions of the deterioration rate parameter $\lambda$ and moment of inertia $I$ at each time step when updating with corrosion measurements only. The prior for $\lambda$ is shown at $t=0$, whereas the prior for $I$ is shown alongside the posterior at each updating step.}
\label{fig:cantilever_results_property}
\end{figure*}

\Cref{fig:cantilever_results_property} shows the posterior distributions of the deterioration rate $\lambda$ at each time step obtained through numerical integration using the observed deterioration data. 

\subsubsection{Updating with Property and Behavior Data}

\begin{figure*}[h]
\centering
\includegraphics[width=\textwidth]{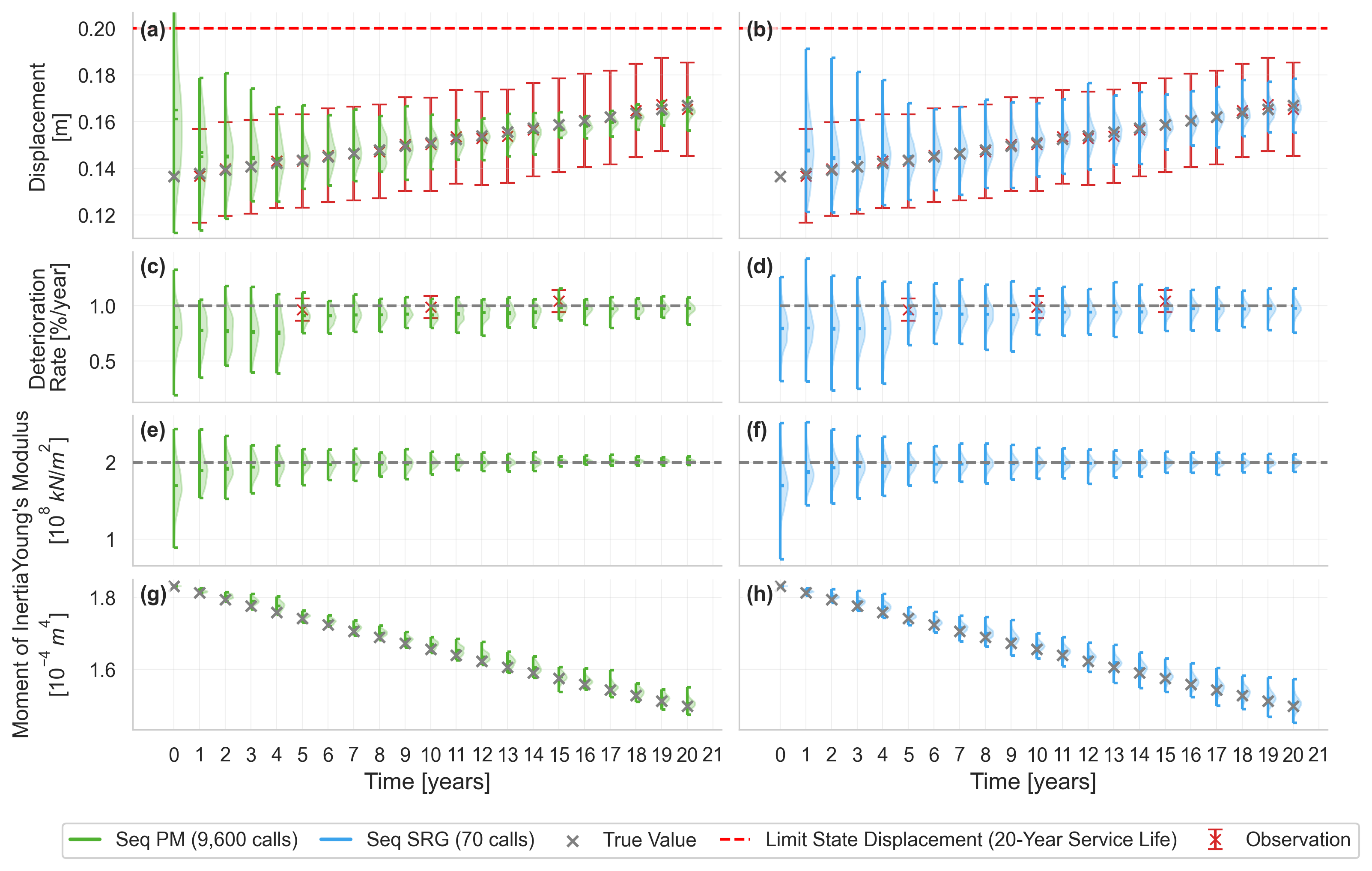}
\caption{Posterior distribution at each time step of the deterioration rate parameter $\lambda$, moment of inertia $I$, Young's Modulus $E$, and predicted displacement using a physics-based model for updating with both corrosion and displacement measurements.}
\label{fig:cantilever_results_behavior}
\end{figure*}

\Cref{fig:cantilever_results_behavior} presents the results obtained when both property and behavior data are used for updating. As above, the deterioration rate is observed in three different years. In addition, displacement measurements are assumed to be available annually. In years where only displacement data are available, only the second step of the updating procedure described in \Cref{subsec:pdt_cont_updating} is performed. In years where both property and behavior data are available, both updating steps are applied. Two scenarios are considered: the left column shows the results for sequential updating with the physics-based model (Seq PM), while the right column shows those obtained with the surrogate model and active learning (Seq SRG).

Panels (a) and (b) of \Cref{fig:cantilever_results_behavior} show the displacement predictions for Seq PM and Seq SRG, respectively. The surrogate predictions closely match those of the physics-based model. The surrogate-based approach requires only $70$ model calls, compared with $9\,600$ for the physics-based approach. The measured annual displacements are also shown. As shown in panels (c) and (d), incorporating behavior data leads to a slight additional reduction in the posterior uncertainty of the deterioration rate compared with updating using property data alone.

The effect of incorporating behavior data is more pronounced for Young's modulus $E$, as shown in panels (e) and (f) of \Cref{fig:cantilever_results_behavior}. Before time step $t=5$, only displacement measurements are available, which provide limited information for identifying $E$ independently. At $t=5$, the additional measurement of the deterioration rate reduces the uncertainty in the deterioration process and, consequently, in the inferred moment of inertia $I$. This improved characterization of $I$ allows the displacement observations to better inform the estimate of $E$, leading to a significant reduction in its posterior uncertainty. This illustrates how combining property and behavior data can reduce uncertainty in coupled state variables. Panels (g) and (h) show the corresponding evolution of the moment of inertia, with clear reductions in uncertainty when deterioration measurements are incorporated.

\subsubsection{Comparison of Computational Strategies}

We investigate four scenarios to compare sequential and batch updating strategies and to assess the performance of surrogate-based approaches relative to physics-based models. The comparison focuses on both posterior accuracy and computational cost. The considered scenarios are: sequential updating with a physics-based model (Seq PM), Sequential updating with a surrogate model and active learning (Seq SRG), Batch updating with a physics-based model (Batch PM), Batch updating with a surrogate model and active learning (Batch SRG).

A reference posterior is obtained by two-dimensional numerical integration over $E$ and $\lambda$, using the physics-based model and the same priors, observations, and measurement-error assumptions as the updating methods.

\begin{figure*}[h]
\centering
\includegraphics[width=\textwidth]{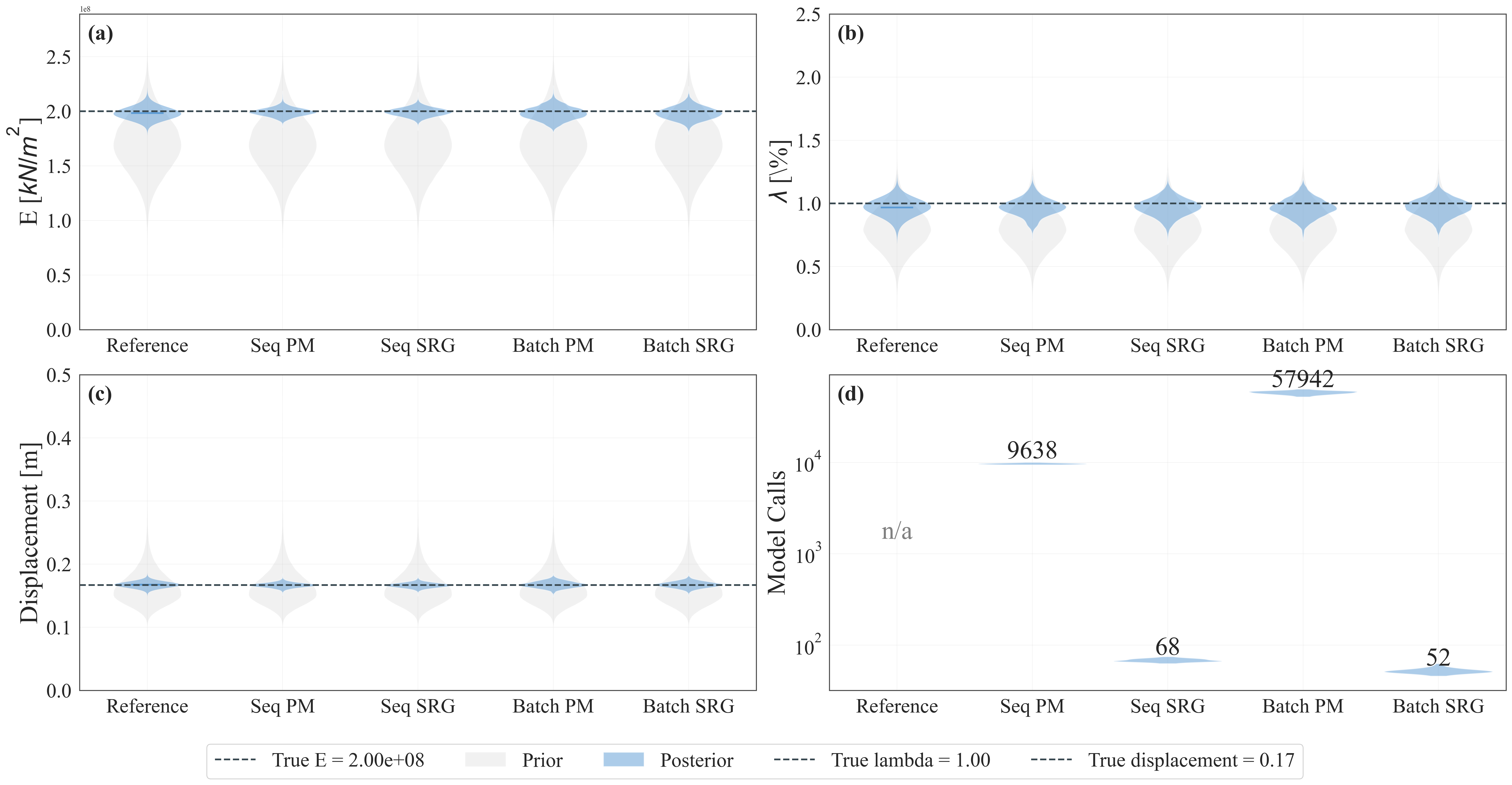}
 \caption{The posterior values obtained at the final time step for $50$ runs for the Young's Modulus $E$, deterioration rate parameter $\lambda$, predicted displacement and number of model calls}
\label{fig:cantilever_results_comparison}
\end{figure*}

The results are illustrated in \Cref{fig:cantilever_results_comparison}, which shows the posterior distributions obtained from $50$ repeated runs. All four approaches yield nearly identical posterior means. Overall, both batch approaches closely match the reference posterior results, but sequential updating underestimates the uncertainty in $E$ relative to the reference standard deviation. The corresponding numerical summaries are provided in \Cref{tab:cantilever_results}.

For the physics-based models, batch updating requires substantially more model evaluations than sequential updating, with an average of $57\,942$ model calls compared to $9\,638$ for the sequential case. This increase arises because all available measurements must be re-evaluated at each time step in the batch formulation.

By contrast, for the surrogate-based approaches, the difference in computational cost between batch and sequential updating is marginal, with only $52$ versus $68$ model calls on average. Although batch updating requires re-evaluating all likelihood terms at each time step, the surrogate model trained in earlier steps is reused more efficiently. This is caused by the active learning stopping criterion, which requires adding at least one training point at each updating step. This highlights a key advantage of surrogate modeling with active learning: batch updating becomes computationally efficient, since surrogate evaluations are inexpensive compared to physics-based model evaluations. 

\begin{table}[htbp]
    \caption{Posterior statistics at the final time step over $50$ runs compared with the reference solution.}
    \label{tab:cantilever_results}
    \centering
    \begin{tabular}{l|cc|cc|cc|cc}
        \hline
        & \multicolumn{2}{c|}{Posterior $E$} & \multicolumn{2}{c|}{Posterior $\lambda$}
        & \multicolumn{2}{c|}{Posterior $\delta_{max}$} & \multicolumn{2}{c}{Model calls} \\
        Method & $\mu$ & $\sigma$ & $\mu$ & $\sigma$ & $\mu$ & $\sigma$ & $\mu$ & $\sigma$ \\
        \hline
        True      & 20.00 & -- & 1.000 & -- & 0.167 & -- & -- & -- \\
        Reference & 19.82 & 0.59 & 0.968 & 0.073 & 0.168 & 0.005 & -- & -- \\
        Batch PM  & 19.82 & 0.60 & 0.965 & 0.073 & 0.167 & 0.005 & 57{,}942 & 2{,}448 \\
        Batch SRG & 19.83 & 0.58 & 0.967 & 0.071 & 0.167 & 0.005 & 52 & 3 \\
        Seq PM    & 19.92 & 0.38 & 0.971 & 0.071 & 0.167 & 0.004 & 9{,}638 & 155 \\
        Seq SRG   & 19.93 & 0.39 & 0.970 & 0.073 & 0.167 & 0.004 & 68 & 3 \\
        \hline
    \end{tabular}
\end{table}

\section{Numerical Investigations: Canal Wall}
\label{sec:example_canal}

\subsection{Setup}

\begin{figure*}[!h]
\centering
\includegraphics[width=0.8\textwidth]{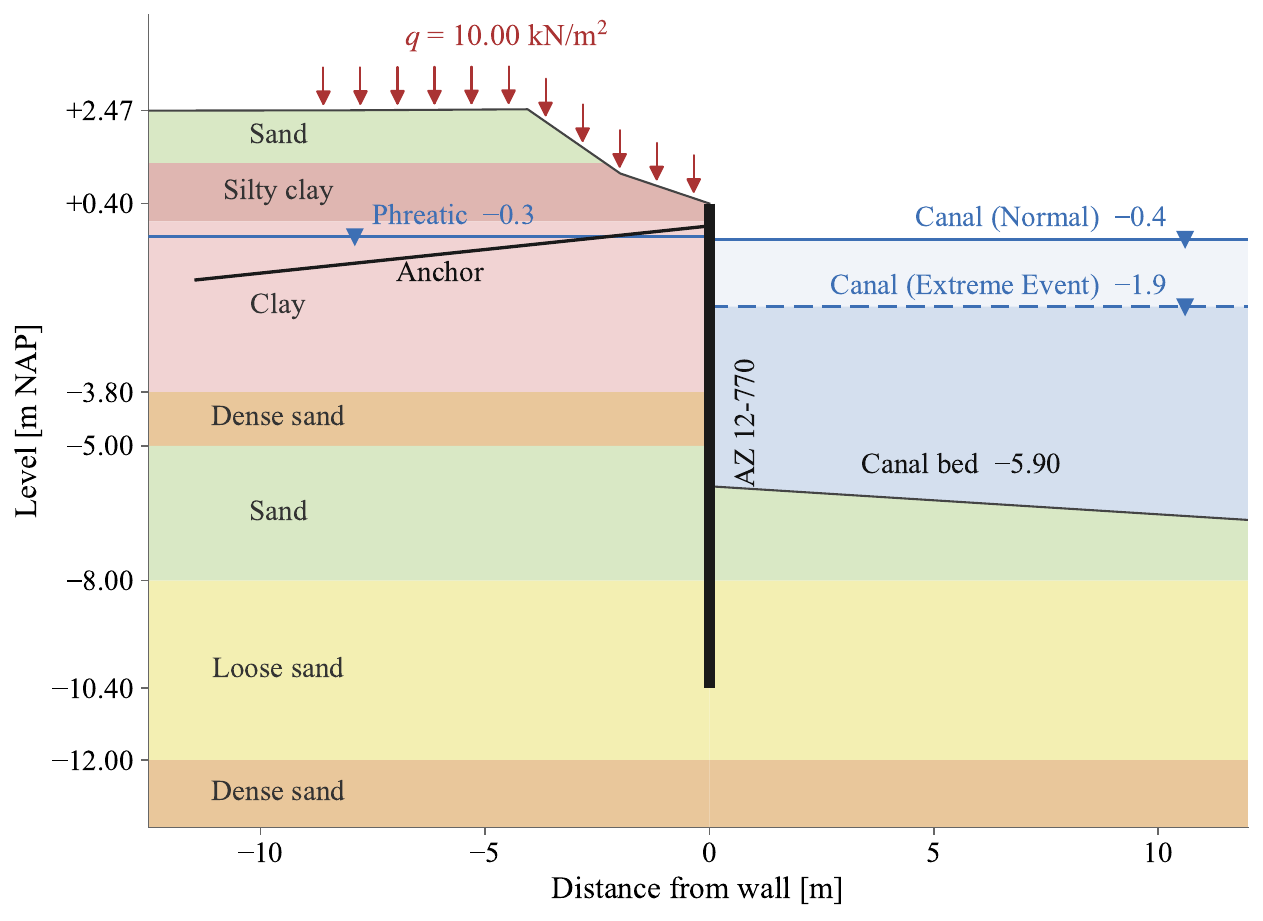}
\caption{Sketch of the cross-section for the given canal problem.}
\label{fig:canal_wall_cross}
\end{figure*}

In the second example, we consider a sheet pile wall structure adjacent to a canal, illustrated in \Cref{fig:canal_wall_cross}. It is inspired by a cross-section of the Amsterdam Rijnkanaal (ARK) located in Utrecht, Netherlands. The objective is to predict the remaining useful life of the canal wall using measurement data collected over time. The detailed problem setup is described below.

As no field data are available for this cross-section, the observations are generated synthetically from an assumed true state. The prior knowledge is deliberately offset from this state to test the ability of the probabilistic digital twin to learn from data. The true soil parameters lie $0.05$--$1.70\,\sigma$ below their prior means. These deviations combine a common site bias of $-0.75\,\sigma$ with an independent layer-specific term drawn from $\mathcal{N}(0,0.5^{2})$ and shared by all parameters within that layer. The true steel yield strength and corrosion parameter each lie $1\,\sigma$ above their respective prior means. The detailed problem setup is described below.

\paragraph{Soil Properties}
    The cross-section consists of seven soil layers with varying thicknesses, composed of different types of sands and clays, extending from $+2.65$ to $-12.00\,\mathrm{m}$ NAP. The soil profile is simplified and we only consider one type of clay and three types of sand (sand, dense sand, and loose sand). Sands are characterized by their friction angle and stiffness. Clays also have cohesion as an additional parameter.  The prior distributions of these parameters are given in \Cref{tab:canal_soil_rvs} and together form the vector $\mathbf{X}_{soil}$.

\begin{table*}[htbp]
    \caption{Prior distributions of the soil properties $\mathbf{X}_{\mathrm{soil}}$.}
    \label{tab:canal_soil_rvs}
    \small
    \centering
    \begin{adjustbox}{width=\textwidth}
    \begin{tabular}{l|l|c|c|c|c|c}
        \hline
        \textbf{Soil type}
        & \textbf{Parameter}
        & \textbf{Symbol}
        & \textbf{Unit}
        & \textbf{Distribution}
        & \textbf{Mean}
        & \shortstack{\textbf{Standard}\\\textbf{deviation}} \\
        \hline
        \multirow{3}{*}{Clay}
        & Friction angle & $\phi_{\mathrm{clay}}$ & ${}^{\circ}$ & Lognormal & $30.77$ & $3.077$ \\
        & Stiffness & $k_{\mathrm{clay}}$ & $\mathrm{kN}/\mathrm{m}^3$ & Lognormal & $30\,000$ & $3\,000$ \\
        & Cohesion & $c_{\mathrm{clay}}$ & $\mathrm{kN}/\mathrm{m}^2$ & Lognormal & $10.03$ & $1.003$ \\
        \hline
        \multirow{2}{*}{Sand}
        & Friction angle & $\phi_{\mathrm{sand}}$ & ${}^{\circ}$ & Lognormal & $27.5$ & $2.75$ \\
        & Stiffness & $k_{\mathrm{sand}}$ & $\mathrm{kN}/\mathrm{m}^3$ & Lognormal & $12\,000$ & $1\,200$ \\
        \hline
        \multirow{2}{*}{Dense sand}
        & Friction angle & $\phi_{\text{sand-dense}}$ & ${}^{\circ}$ & Lognormal & $33.8$ & $3.38$ \\
        & Stiffness & $k_{\text{sand-dense}}$ & $\mathrm{kN}/\mathrm{m}^3$ & Lognormal & $30\,000$ & $3\,000$ \\
        \hline
        \multirow{2}{*}{Loose sand}
        & Friction angle & $\phi_{\text{sand-loose}}$ & ${}^{\circ}$ & Lognormal & $30$ & $3.0$ \\
        & Stiffness & $k_{\text{sand-loose}}$ & $\mathrm{kN}/\mathrm{m}^3$ & Lognormal & $12\,000$ & $1\,200$ \\
        \hline
    \end{tabular}
    \end{adjustbox}
\end{table*}

\paragraph{Wall Properties}
    The wall consists of \textit{AZ 12-770} sheet pile profiles. The structural behavior is described by two parameters: the steel yield strength, $f_{y,wall}$, and the section modulus, $W_{wall}$, which decreases over time due to corrosion. The wall is reinforced with an anchor; however, anchor corrosion is neglected for simplicity. The prior parameters are listed in \Cref{tab:canal_wall_rvs}.

\paragraph{Corrosion Model}
    Corrosion is modeled as a cumulative reduction in the sheet pile thickness, which directly affects the structural capacity by reducing its section modulus $W_{wall}$. Following \citep{g_van_der_krogt_time-dependent_2022,TNO2023richtlijn}, a corrosion model based on the Dutch standard NEN 6766 \citep{NEN6766:2021ontwerp} is adopted for sheet piles in fresh water conditions. The thickness loss at time $t$ is given by
    
    \begin{equation}
    \label{eq:corrosion}
        C_r(t)= A_{CR} \cdot t^{0.744}\\
    \end{equation}
    where $A_{CR}$ is the corrosion rate parameter for soil conditions of type~1 in fresh water, specified in \Cref{tab:canal_wall_rvs}.

\paragraph{Loads}
    A fixed uniform load $V$ is applied on the embankment side of the wall, representing a nearby structure. In addition, the wall is subjected to earth and hydrostatic pressures induced by the water levels on both sides.

\begin{figure*}[h]
    \centering
    \begin{subfigure}[]
        \centering
        \includegraphics[width=0.59\textwidth]{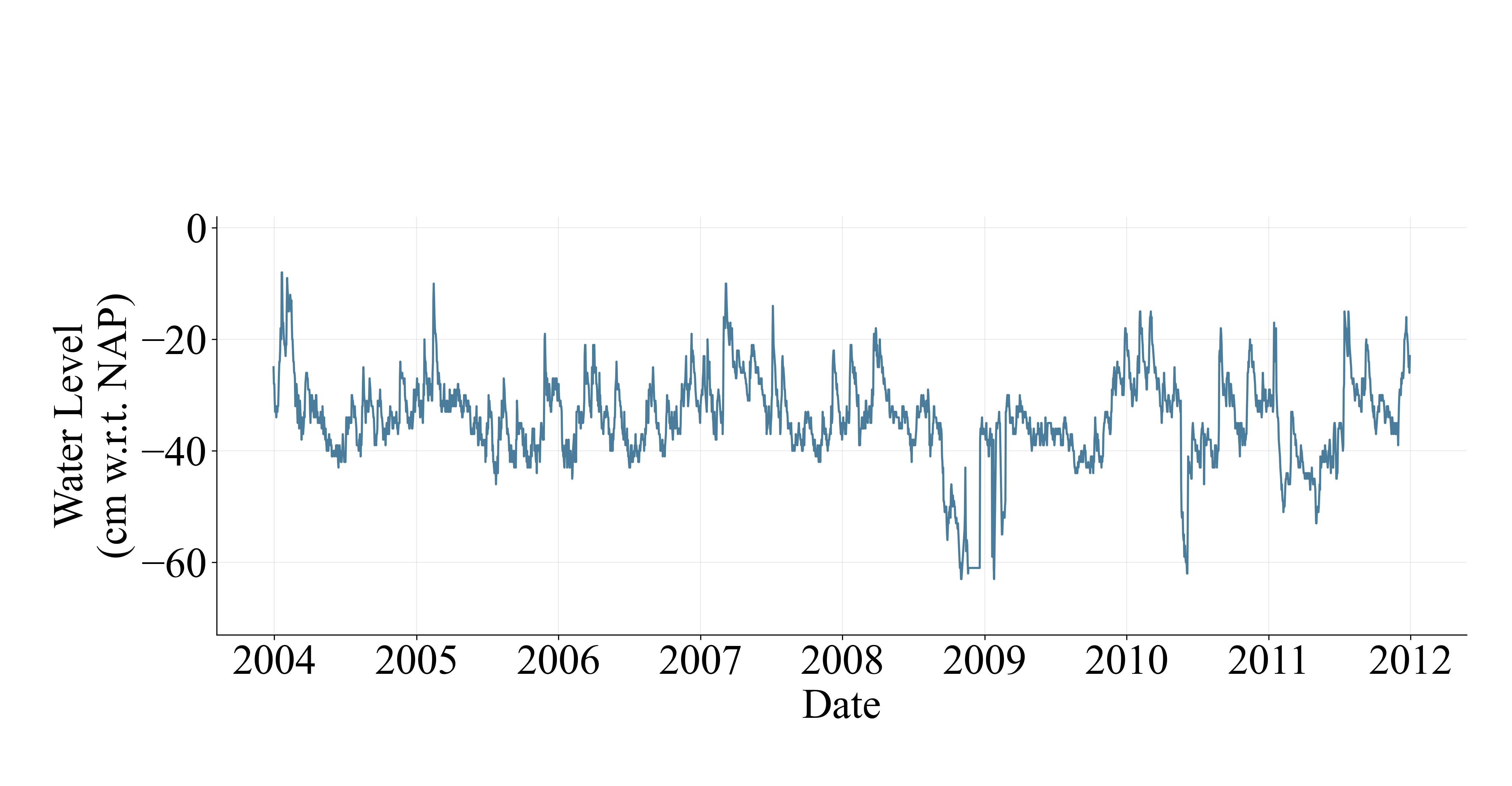}
        \label{fig:canal_gw_timeseries}
    \end{subfigure}
    % \hfill
    \begin{subfigure}[]
        \centering
        \includegraphics[width=0.38\textwidth]{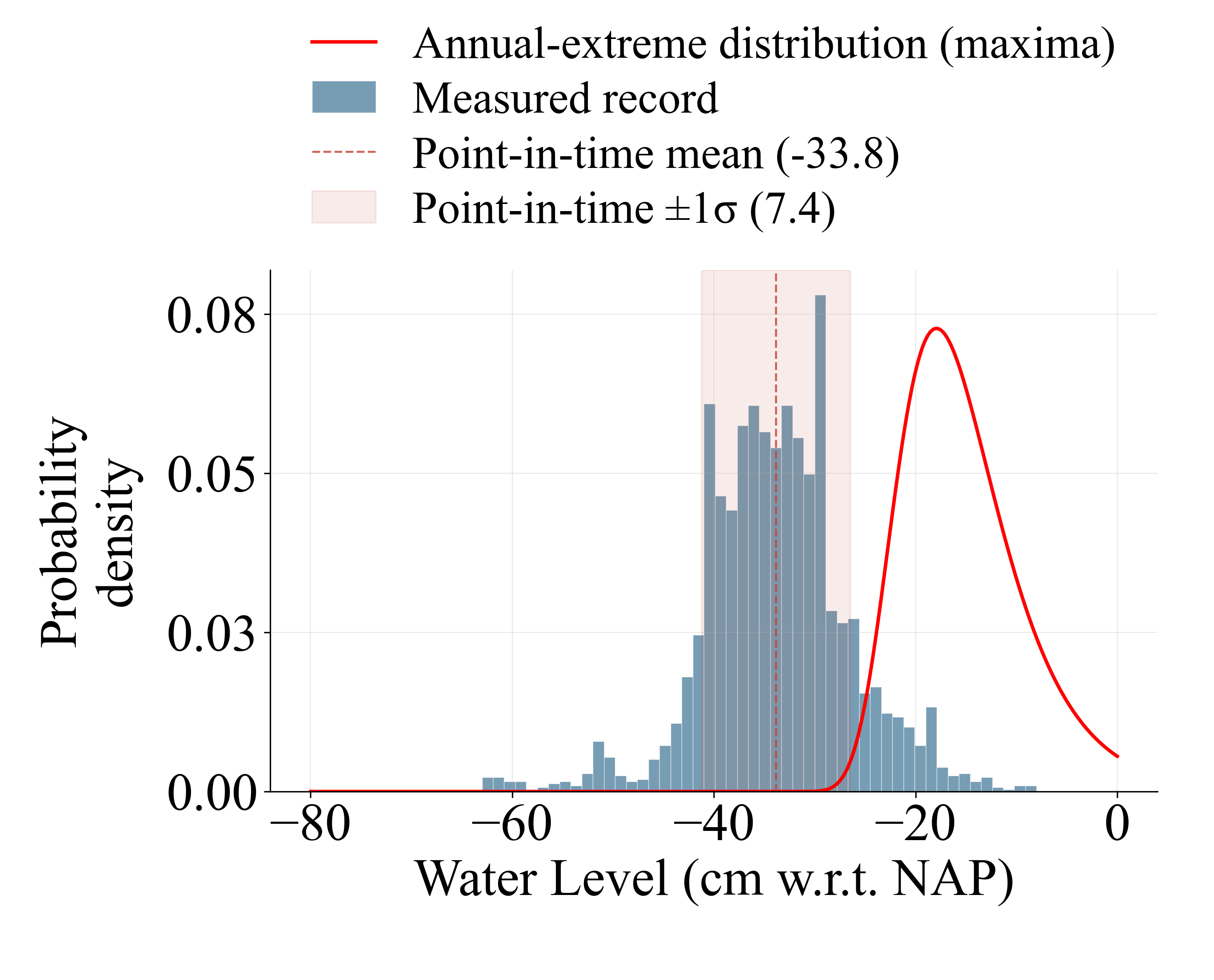}
        \label{fig:canal_gw_analysis}
    \end{subfigure}
    \caption{(a) Phreatic water level timeseries data from $2003-2012$ (b) Histogram of the data with the Gumbel distribution of the yearly maxima learned from the data.}
    \label{fig:phreatic_wl}
    \end{figure*}

\paragraph{Phreatic Water Level}
    The phreatic water level is measured within the soil and contributes to the loading on the landward side of the wall. Historical data on the phreatic water level are available in public Dutch databases \citep{grondwatertools2026}. We selected a measurement station with data from the years $2003$ up to $2012$ in close proximity to the considered cross section, as shown in \Cref{fig:phreatic_wl}. These data are used to estimate the statistical properties of the water level, as summarized in \Cref{tab:canal_wall_rvs}.

\begin{figure*}[h]
    \centering
    \begin{subfigure}[]
        \centering
        \includegraphics[width=0.59\textwidth]{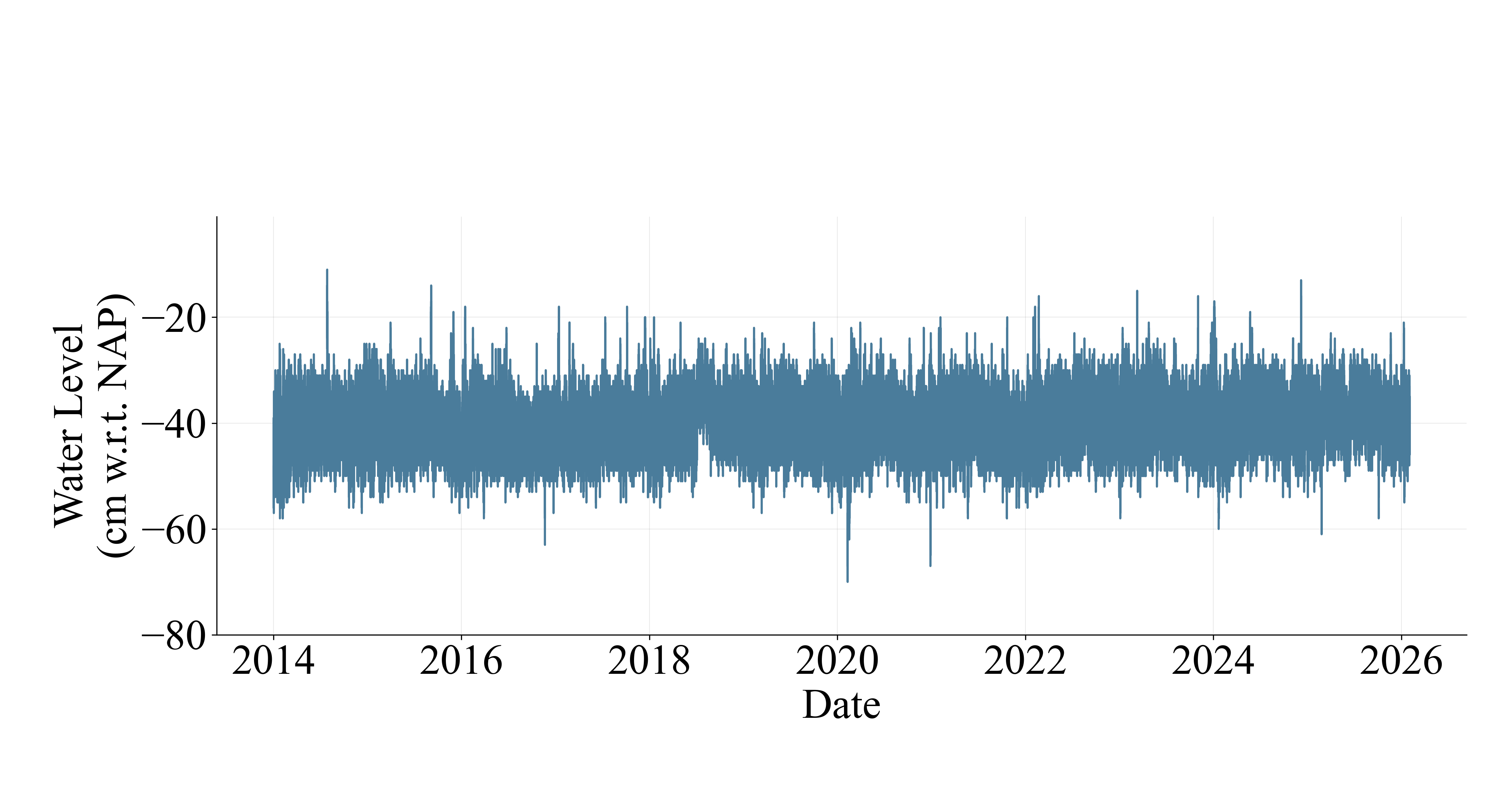}
        \label{fig:canal_wl_timeseries}
    \end{subfigure}
    % \hfill
    \begin{subfigure}[]
        \centering
        \includegraphics[width=0.38\textwidth]{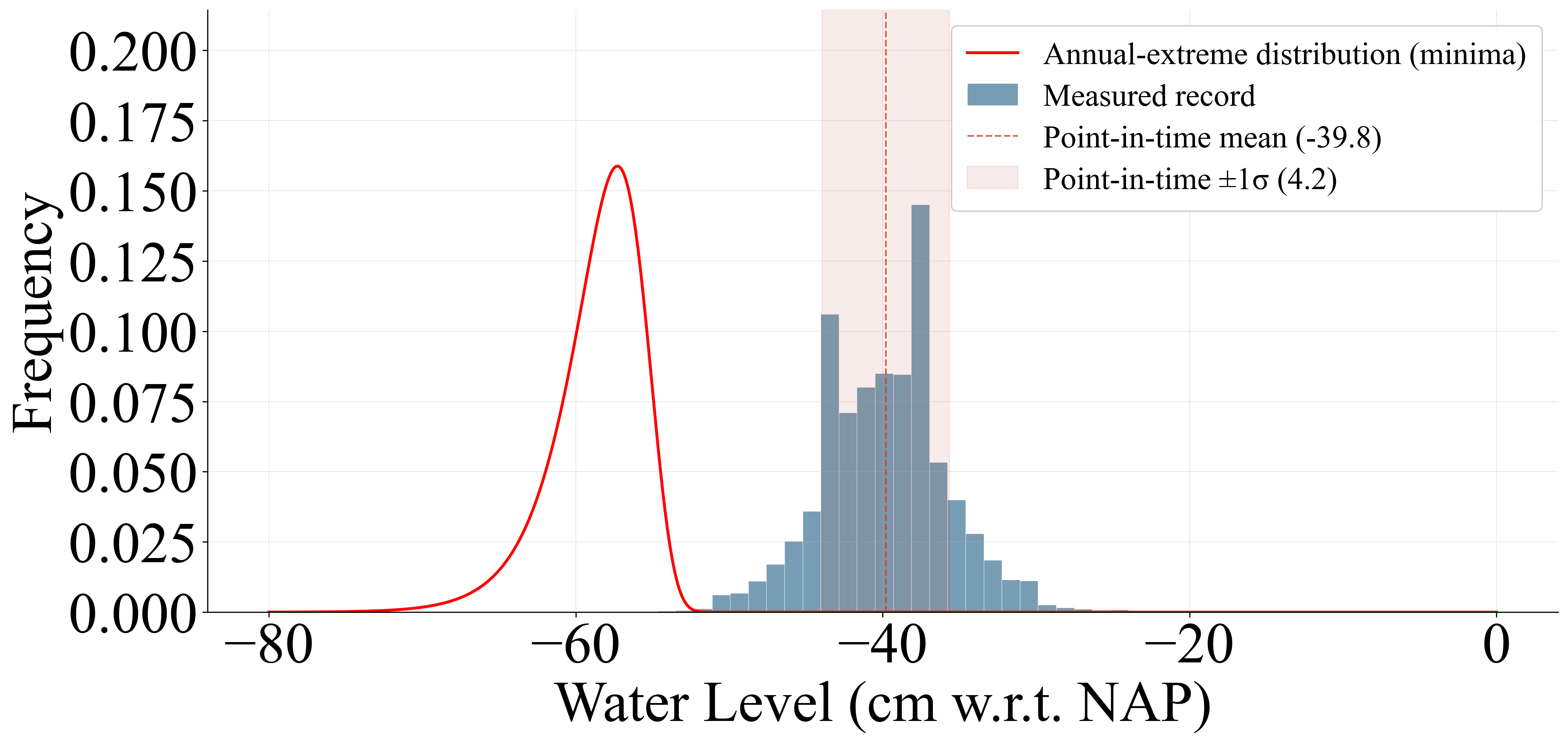}
        \label{fig:canal_wl_analysis}
    \end{subfigure}
    \caption{(a) Canal water level timeseries data from $2014$ to $2026$ (b) Histogram of the data with an extreme distribution of the yearly minima.}
    \label{fig:canal_wl}
    \end{figure*}

\paragraph{Canal Water Level}
    The canal water level provides a stabilizing counter-pressure on the canal side of the wall. The water level is actively regulated, and data can be accessed online from \citet{waterinfoRWS2026}. In \Cref{fig:canal_wl}, we illustrate the available time series data from the years $2014$ to $2026$. These data are used to estimate the statistical properties of the canal water level, as reported in \Cref{tab:canal_wall_rvs}.

\paragraph{Basin Depth}
    The canal bed slopes from $-5.90\,\mathrm{m}$ NAP at the wall to $-7.10\,\mathrm{m}$ NAP in the far field. For simplicity, erosion or deposition of the canal bed is not considered. If relevant, this could be included as an additional random variable.

\begin{table*}[htbp]
    \caption{Sheet pile and environmental parameters.}
    \label{tab:canal_wall_rvs}
    \centering
    \begin{adjustbox}{width=\textwidth}
    \begin{tabular}{l|c|c|c|c|c}
        \hline
        \textbf{Parameter}
        & \textbf{Symbol}
        & \textbf{Unit}
        & \textbf{Distribution}
        & \shortstack{\textbf{Mean/}\\ \textbf{Value}}
        & \shortstack{\textbf{Standard}\\\textbf{deviation}}\\
        \hline
        Initial sheet pile thickness
        & $d_0$
        & $\mathrm{mm}$
        & Deterministic
        & $8.5$
        & -- \\

        Initial section modulus
        & $W_{0,\mathrm{wall}}$
        & $\mathrm{cm}^3/\mathrm{m}$
        & Normal
        & $1\,245$
        & $124.5$ \\

        Steel yield strength (S355GP)
        & $f_{y,\mathrm{wall}}$
        & $\mathrm{MPa}$
        & Normal
        & $355$
        & $35.5$ \\

        Corrosion parameter
        & $A_{\mathrm{CR}}$
        & $\mathrm{mm}\,\mathrm{year}^{-0.744}$
        & Normal
        & $0.3226$
        & $0.1000$ \\
        \hline
        Phreatic water level, point-in-time
        & $H_{\mathrm{ph}}$
        & $\mathrm{m}$
        & Normal
        & $-0.34$
        & $0.074$ \\

        Phreatic water level, annual maximum
        & $H_{\mathrm{ph}}^{\max}$
        & $\mathrm{m}$
        & Gumbel (max)
        & $-0.14$
        & $0.043$ \\

        Canal water level, point-in-time
        & $H_{\mathrm{can}}$
        & $\mathrm{m}$
        & Normal
        & $-0.40$
        & $0.042$ \\

        Canal water level, annual minimum
        & $H_{\mathrm{can}}^{\min}$
        & $\mathrm{m}$
        & Gumbel (min)
        & $-0.59$
        & $0.040$ \\

        Surcharge load
        & $V$
        & $\mathrm{kN}/\mathrm{m}^2$
        & Uniform
        & $10$
        & $1.0$ \\

        Basin depth
        & $D$
        & $\mathrm{m}$
        & Deterministic
        & $-7.10$ to $-5.90$
        & -- \\
        \hline
    \end{tabular}
    \end{adjustbox}
\end{table*}
    
\paragraph{Prediction Models}
    To predict bending moments in the wall $M_{wall}$, we use Deltares D-SheetPiling \citep{Deltares_DSheetPiling_2014}, an engineering software for the analysis of sheet pile walls based on a beam–spring formulation. The model accounts for soil–structure interaction, layered soil profiles, groundwater levels, and external loads, and is commonly applied for reliability assessment in engineering practice. For this example, its runtime is approximately $10$ seconds per model run on a standard laptop, which makes sampling-based Bayesian updating computationally expensive when a large number of model evaluations is required. A GPR surrogate model with active learning is used to approximate the original model and reduce computation time.
        
\paragraph{Limit State}
    A single limit state is considered, corresponding to yielding of the steel due to bending. Following \citet{g_van_der_krogt_time-dependent_2022}, the limit-state function is defined as
    \begin{equation}
    \label{eq:canal_lsf}
        g_{wall}(\mathbf{X},t)=f_{y,wall}-\theta_M\cdot \frac{\max M_{wall}(\mathbf{X})}{W_{wall}(\mathbf{X},t)},
    \end{equation}
      where $\theta_M$ represents model uncertainty in the bending moment prediction. For simplicity, model uncertainty is neglected in this work by setting $\theta_M=1$.

\paragraph{Load Combinations}
Since the limit state function depends on the time-varying water levels, we need to address the resulting load-combination problem. To this end, we adopt Turkstra’s rule \citep{turkstra_load_1980}, which approximates the maximum load by considering in turn each time-variable parameter as the leading variable, represented by its extreme-value distribution over the chosen reference period, while the remaining loads are represented by their point-in-time distributions. The maximum combined load over the reference time period is then approximated by the maximum among all cases.

In our application, each water level is considered as the leading load once. In case~(a), the canal water level is leading and is described by the Gumbel distribution fitted to the yearly minima, while the phreatic level takes its point-in-time distribution. In case~(b), the phreatic level is leading and is described by the Gumbel distribution fitted to the yearly maxima, while the canal level takes its point-in-time distribution. 
As a further approximation to Turkstra's rule, we evaluate the failure probability for the two load cases separately at every time step and take the failure probability as the maximum value, $P_f(t)=\max_{k\in\{a,b\}}P_f^{(k)}(t)$, where $P_f^{(k)}(t)$ denotes the failure probability obtained with the water levels of case~$k$. This results in a lower-bound estimate, which, however, is within a few percent of the true result in most cases.
%upper bound?//couldn't get api access to redo correct way

\paragraph{Measurement Data}
   We consider three types of available measurements: corrosion measurements, site investigation data, and survival information.  As noted above, the observations
  are generated synthetically.

\begin{figure*}[!h]
\centering
\includegraphics[width=0.6\textwidth]{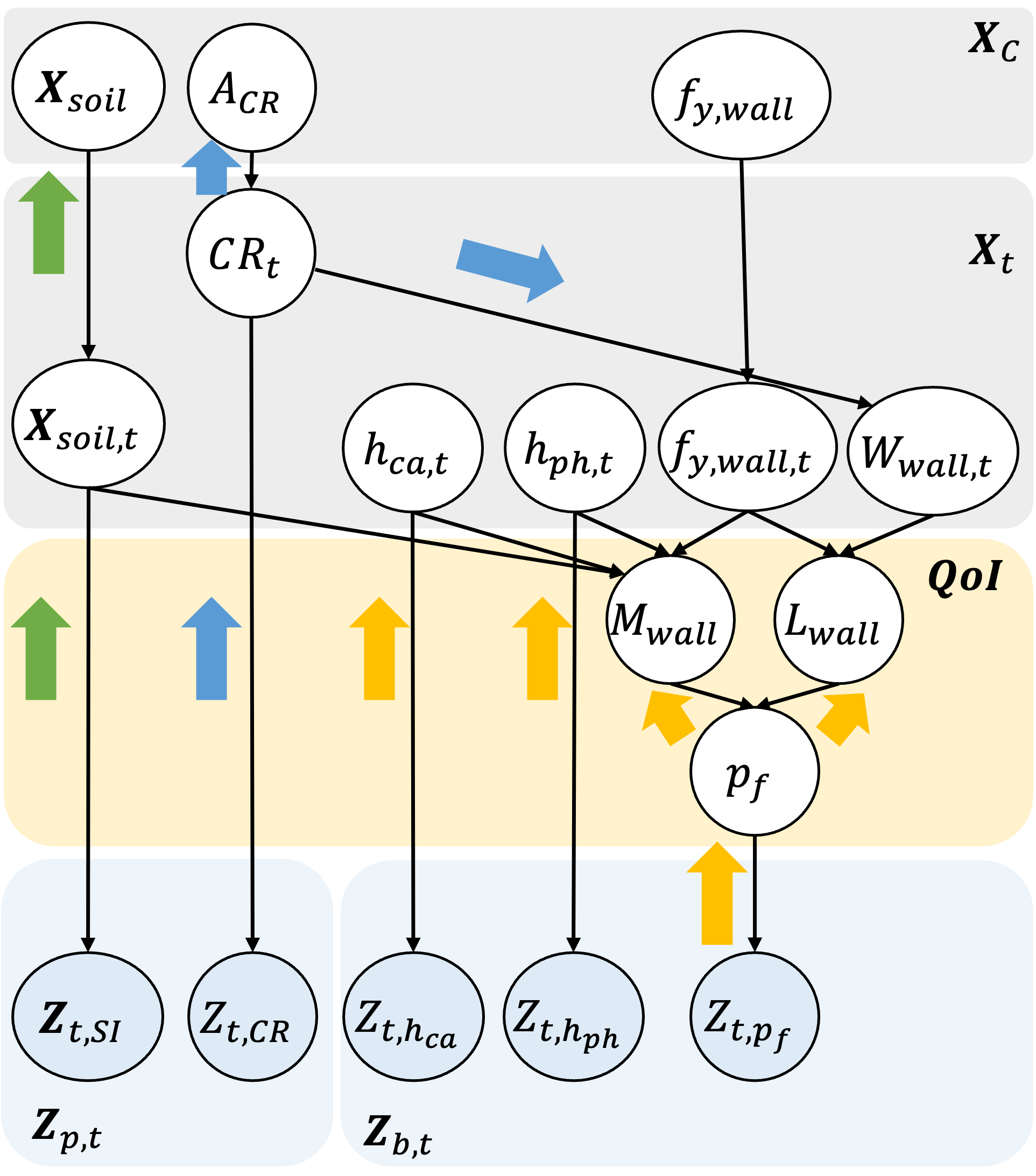}
\caption{Bayesian network of the canal wall example problem.}
\label{fig:canal_wall_network}
\end{figure*}

    \begin{enumerate}
    \item Measured corrosion.\\
    Every five years, the canal wall is inspected for corrosion. The measured quantity is the thickness loss expressed as a fraction of the initial sheet pile thickness $d_0=8.5\,\mathrm{mm}$. This information is classified as \textit{property data}, as it directly informs the corrosion parameter $A_{\mathrm{CR}}$ through the corrosion model (blue arrows in \Cref{fig:canal_wall_network}). For an observed value $z_{\mathrm{CR},t}$, the likelihood is
    \begin{equation}
        L_{{\mathrm{CR,t}}}(A_{\mathrm{CR}})
        =
        f_{\varepsilon_{\mathrm{CR}}}\!\left(
            z_{\mathrm{CR},t}
            -
            \frac{A_{\mathrm{CR}}\,t^{0.744}}{d_0}
        \right),
    \end{equation}
    where $f_{\varepsilon_{\mathrm{CR}}}$ is the PDF of the additive measurement error, assumed to follow a zero-mean normal distribution:
    \begin{equation}
        \varepsilon_{\mathrm{CR}}
        \sim
        \mathcal{N}\!\left(0,\sigma_{\mathrm{CR}}\right).
    \end{equation}
    The standard deviation $\sigma_{\mathrm{CR}}=0.05$ corresponds to $5\%$ of the initial thickness, or approximately $0.43\,\mathrm{mm}$.

\item Site investigation.\\
During the considered lifetime of the canal wall, one in-situ investigation is assumed to provide observations of all parameters in $\mathbf{X}_{\mathrm{soil}}$, as indicated by the green arrows in \Cref{fig:canal_wall_network}. For an observed value $z_{i,t}$ of a soil parameter $X_i$, the likelihood is
\begin{equation}
    L_{SI,i}(x_i)
    =
    f_{\varepsilon_{i,t}}\!\left(z_{i,t}-x_i\right),
\end{equation}
where $f_{\varepsilon_{t,i}}$ is the PDF of the additive measurement error, assumed to follow a zero-mean normal distribution, whose standard deviation $\sigma_{i}$ is set equal to $5\%$ of the parameter's true values:
\begin{equation}
    \varepsilon_{t,i}
    \sim
    \mathcal{N}\!\left(0,\sigma_{i}^2\right).
\end{equation}

Synthetic measurements are generated by adding independent, zero-mean Gaussian noise with a standard deviation of $2\%$ of the true value of the parameter.

\item Water levels and survival information.\\
For the survival update, we assume that the phreatic and canal water levels are measured,
\begin{equation}
    \mathbf{Z}_{\mathbf{h},t}
    =
    \left[
    z_{h_{\mathrm{phreatic}},t},
    z_{h_{\mathrm{canal}},t}
    \right].
\end{equation}

Annual water-level measurements are generated synthetically, assuming Gaussian measurement errors with standard deviations of $0.015\,\mathrm{m}$ for the canal level and $0.020\,\mathrm{m}$ for the phreatic level. To include a rare loading condition, the canal level is lowered to $-2.00\,\mathrm{m}$ NAP for maintenance at $t=13$. Survival of this event is the most informative behavior observation in the record. Unlike the reliability evaluation, which uses the load combinations defined above, the survival update conditions directly on the measured levels. Each level is represented in the limit state function by a normal distribution centered on its measured value, with the corresponding measurement-error standard deviation. Together with observed survival, these measurements provide information on the system capacity, as indicated by the yellow arrows in \Cref{fig:canal_wall_network}.

Utilizing the limit state function $g_{\mathrm{wall}}$ of \Cref{eq:canal_lsf}, the observed survival event at time $t$ is defined as
\begin{equation}
    \mathcal{S}_t
    =
    \left\{
    g_{\mathrm{wall}}(\mathbf{Y}_t,\mathbf{z}_{\mathbf{h},t},t)>0
    \right\},
\end{equation}
in which $\mathbf{Y}_t$ contains all random variables in $\mathbf{X}_t$ other than the water levels.
This corresponds to inequality information in structural reliability \citep{straub2011reliability}. The corresponding likelihood is defined by the indicator function
\begin{equation}
    L_{\mathcal{S}_t}(\mathbf{x})
    =
    I\!\left[
    g_{\mathrm{wall}}(\mathbf{y}_t,\mathbf{z}_{\mathbf{h},t},t)>0
    \right].
\end{equation}

\end{enumerate}

\paragraph{Updating procedure}
    Updating at each time step proceeds in two stages. First, the property data are used to update $A_{CR}$ based on the corrosion measurements $z_{CR,t}$ and, separately, $\mathbf{X}_{soil}$ based on the site investigation data $\mathbf{z}_{SI,t}$ using numerical integration. These updates concern different subsets of $\mathbf{X}_t$ and can be performed in either order. Second, $\mathbf{X}_t$ is updated using the observed survival and measured water levels $\mathbf{z}_{\mathbf{h},t}$ through aBUS. This stage can follow either a sequential or a batch updating strategy.

\subsection{Results}

\begin{figure*}[h!]
    \centering
    \includegraphics[width=\textwidth]{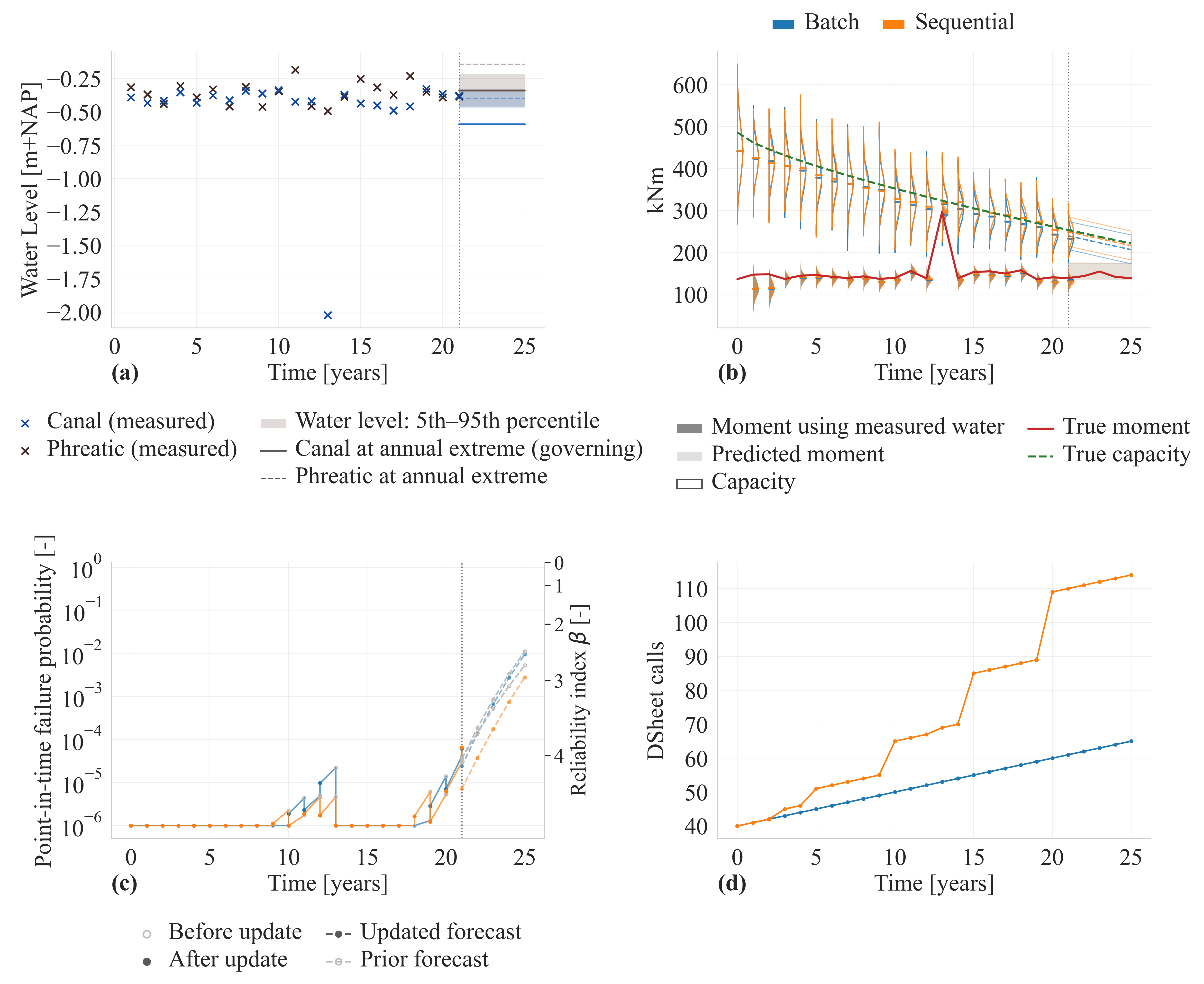}
    \caption{Evolution of the canal-wall digital twin over time: measured and predicted canal and phreatic water levels; predicted bending moment and capacity compared with the true values; point-in-time failure probability and reliability index; and cumulative number of D-SheetPiling model evaluations.}
    \label{fig:canal_dt}
\end{figure*}

\begin{figure*}[h!]
\centering
\includegraphics[width=\textwidth]{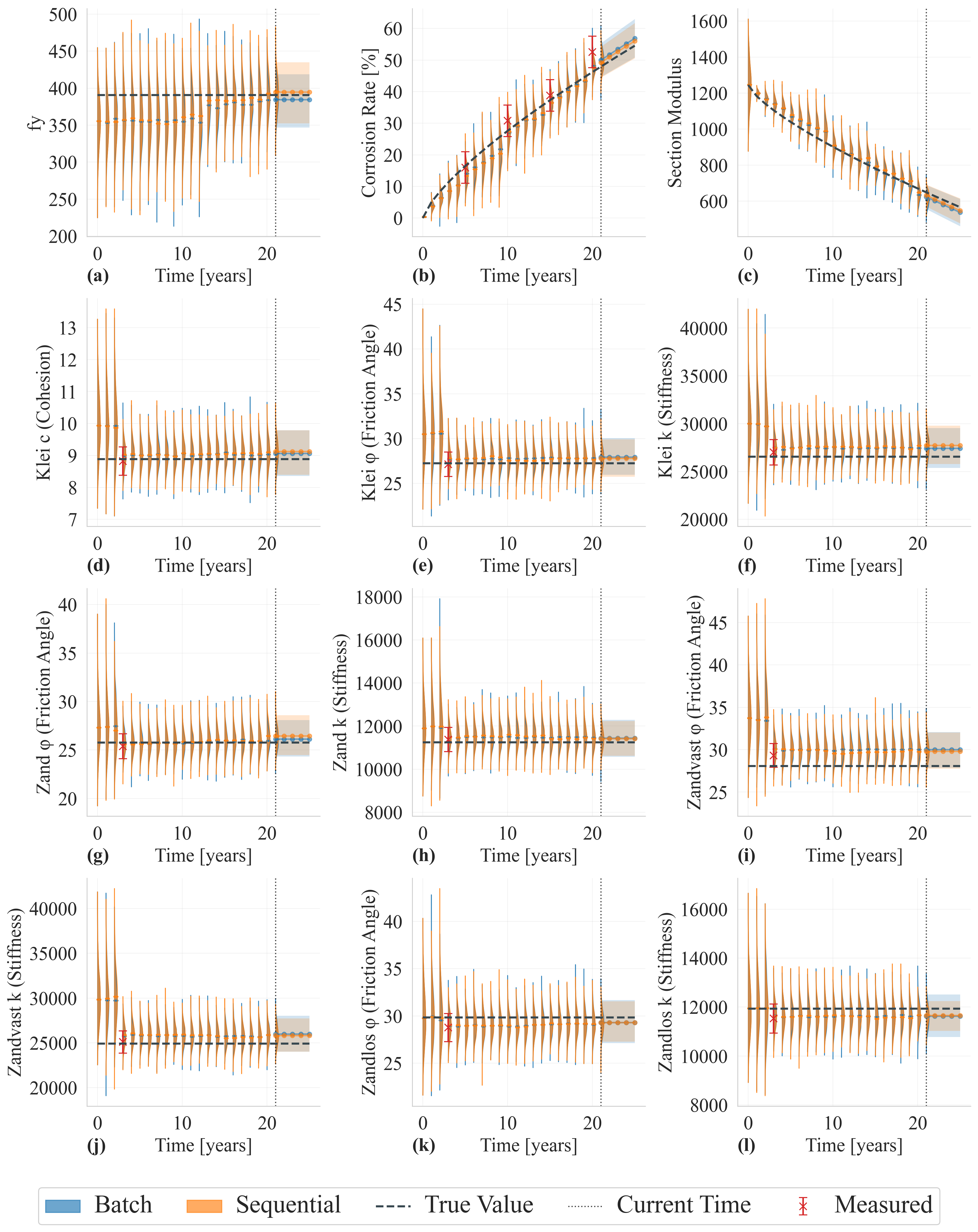}
\caption{Evolution of the wall and soil-parameter distributions over time as new observations are incorporated.}
\label{fig:canal_parameters_dt}
\end{figure*}

\Cref{fig:canal_dt} summarizes input data and key results and \Cref{fig:canal_parameters_dt} provides detailed insights into the parameter updates in function of time. The predicted bending moments in \Cref{fig:canal_dt}b show good agreement with the true response, without any noticeable difference between sequential and batch updating strategies. 

The property data updates have a clear effect on the posterior state. First, the site investigation at $t=3$ significantly reduces the uncertainty in the soil parameters, as shown in \Cref{fig:canal_parameters_dt}, and improves the prediction of the bending moment in \Cref{fig:canal_dt}b. Second, the corrosion measurements at $t=[5,10,15,20]$ directly update the corrosion parameter. After updating the corrosion for the first measurement at $t=10$ the uncertainty in this parameter stays similar over time, as shown in \Cref{fig:canal_parameters_dt}c.

Behavior data are incorporated through survival observations under the measured water levels at each time step, as shown in \Cref{fig:canal_dt}a. These observations are most informative when the measured water levels correspond to extreme loading conditions. For example, the survived load case at $t=13$ increases the inferred resistance of the system, reflected in the posterior distribution of the steel yield strength $f_y$ in \Cref{fig:canal_parameters_dt}. This results in a more accurate representation of the canal-wall resistance and a reduction in the expected probability of failure.

As shown in \Cref{fig:canal_dt}d, starting from an initial surrogate with $40$ samples, the refinement procedure converges to $114$ and $65$ training points for the sequential and batch updating strategies, respectively. As in the example of \Cref{sec:example_cantilever}, the larger number of model calls for the sequential strategy is mainly caused by the additional training point required at each updating step to check for convergence. This effect is particularly visible after the corrosion measurements at $t=10$, $15$, and $20$, where new property data trigger a new behavior-updating loop.

For this case study, the batch and sequential updating strategies lead to similar posterior distributions and reliability estimates, and the batch updating leads to slightly lower computational cost. The results show that combining property data, behavior data, and survival information provides a more complete representation of the canal-wall state and enables the point-in-time probability of failure to be updated over time. Furthermore, by combining the deterioration model with a probabilistic load model for the water levels, the digital twin can forecast the expected reliability beyond the current observation period. This demonstrates the potential of the proposed framework for scenario analysis and reliability-informed decision support.

\section{Concluding remarks}
\label{sec:concluding_remarks} 

This study proposed a framework for multi-source Bayesian updating within a probabilistic digital twin, with application to geotechnical systems. By distinguishing between property data and behavior data, heterogeneous observations can be integrated within a single probabilistic state representation while updating only the relevant parts of the digital state. The numerical examples demonstrated the complementary information provided by the two types of data. Property data, such as deterioration measurements and site investigation data, directly reduce uncertainty in the associated parameters, whereas behavior data, such as displacement measurements and survival observations, provide indirect information through the prediction model and can thereby update coupled parameters that influence the system response. 
%The canal wall example further showed that survival under an extreme water-level event provides information on the system resistance and affects the reliability estimate. 

%The examples illustrated the value of incorporating both equality observations and inequality information into infrastructure digital twins and enables a more complete and continuously updated representation of the physical asset state and its reliability. 

We proposed and investigated different updating strategies. Both sequential and batch updating produced similar posterior distributions in the numerical examples. For physics-based models, sequential updating is computationally more efficient because it avoids re-evaluating the complete history of behavior data. When surrogate models are used, however, batch updating is computationally feasible at almost no additional cost because surrogate evaluations are inexpensive. 
The preferred updating strategy therefore depends on the computational cost of the prediction model and on the quality of the posterior approximation employed between sequential updates. 

We also proposed and implemented an active learning strategy for sequential learning of the surrogate within the probabilistic digital twin. As the digital state evolves through Bayesian updating, the region of the input space relevant to posterior inference may shift. The proposed approach allows the surrogate model to evolve jointly with the digital twin by actively refining it in regions that are important for the current posterior. In the numerical investigations, this substantially reduced the number of expensive model evaluations while yielding posterior estimates comparable to those obtained through physics-based updating. The results also showed that the active-learning stopping criterion influences the relative computational cost of sequential and batch updating. Requiring at least one additional training point at each updating step may introduce unnecessary model evaluations in sequential updating when the surrogate is already sufficiently accurate. Although the stopping criterion adopted in this study performed better than commonly used alternatives, such as the $\min U$ criterion, in the examples considered, less conservative stopping criteria may provide further computational savings. 

Overall, the proposed combination of multi-source Bayesian updating and adaptive surrogate modeling provides a scalable framework for probabilistic digital twins of infrastructure systems. For larger systems, such digital twins could be deployed at representative locations to support monitoring and reliability assessment while limiting the number of computationally expensive model evaluations. Future work should extend the framework toward decision optimization and value-of-information analysis.

\paragraph{Data and code availability}
The measurement data used in the numerical investigations were generated synthetically. The code, configuration files, and instructions for generating the synthetic measurements and reproducing the cantilever beam numerical investigation are available at \url{https://github.com/dafyddctrb/pdt-beam-active-learning}.

\paragraph{Competing interest} The authors declare no competing interests exist.

\paragraph{Funding Statement}
This research is supported by the TUM Georg Nemetschek Institute Artificial Intelligence for the Built World.

\paragraph{Acknowledgement}
We thank Eleni Smyrniou for facilitating access to D-Sheet Piling and for her patience and continued support in maintaining the API access, and Bram van den Eijnden, Mark van der Krogt, and Anton van der Meer for helpful discussions during the development of the canal wall numerical investigation.

\bibliographystyle{elsarticle-harv}
\bibliography{bibliography.bib}

\end{document}